\documentclass[aps,prb,amsmath,amssymb,superscriptaddress,floatfix,showpacs,twocolumn]{revtex4}

\usepackage[pdftex]{graphicx} 
\usepackage{epstopdf} 
\usepackage{subfigure}
\usepackage[usenames,dvipsnames]{color}
\usepackage{bm}

\usepackage{wasysym}
\usepackage{slashed}
\usepackage{tikz}

\newcommand{\osum}{ 
  \mathop{
    \mathchoice
      {\buildosum{\displaystyle}{0.1}}
      {\buildosum{\textstyle}{0.075}}
      {\buildosum{\scriptstyle}{0.075}}
      {\buildosum{\scriptscriptstyle}{0.075}}
  }\displaylimits 
}

\newcommand\buildosum[2]{%
  \begin{tikzpicture}[baseline=(char.base), inner sep=0, outer sep=0]
    \draw (-0.3ex,0) circle (#2);
    \node (char) at (0,0) {$#1\sum$};
  \end{tikzpicture}%
}

\usepackage{hyperref}

\begin{document}

\title[Quantum Spin Ice]{Quantum Spin Ice: A Search for Gapless Quantum Spin Liquids in Pyrochlore Magnets}

\author{M. J. P. Gingras \footnote{gingras@uwaterloo.ca}}
\address{Department of Physics and Astronomy, University of Waterloo, 200 University Avenue West, Waterloo, ON, N2L 3G1, Canada.}
\address{Perimeter Institute for Theoretical Physics, 31 Caroline North, Waterloo, Ontario, N2L-2Y5, Canada}
\address{Canadian Institute for Advanced Research, Toronto, Ontario, Canada, M5G 1Z8}
\author{P. A. McClarty}
\address{Max Planck Institute for the Physics of Complex Systems, N\"{o}thnitzer Strasse 38, Dresden, 01187, Germany.}
\address{ISIS Neutron and Muon Source, STFC, Rutherford Appleton Laboratory, Harwell Oxford, Didcot OX11 0QX, UK.}

\begin{abstract}
The spin ice materials, including Ho$_{2}$Ti$_{2}$O$_{7}$ and Dy$_{2}$Ti$_{2}$O$_{7}$, are rare earth pyrochlore magnets which, at low temperatures, enter a constrained paramagnetic state with an emergent gauge freedom. Remarkably, the spin ices provide one of very few experimentally realised examples of fractionalization because their elementary excitations can be regarded as magnetic monopoles and, over some temperature range, the spin ice materials are best described as liquids of these emergent charges.  In the presence of quantum fluctuations, one can obtain, in principle, a quantum spin liquid descended from the classical spin ice state characterised by emergent photon-like excitations. Whereas in classical spin ices the excitations are akin to electrostatic charges, in the quantum spin liquid these charges interact through a dynamic and emergent electromagnetic field. In this review, we describe the latest developments in the study of such a quantum spin ice, focussing on the spin liquid phenomenology and the kinds of materials where such a phase might be found.
\end{abstract}

\maketitle

\tableofcontents

\section{Introduction}

Leaving aside molecular magnets, magnetic chains and layered magnets, there are many thousands  of magnetic materials known to us. These typically exhibit a low temperature phase with some long-range ordered magnetic structure which, no matter how complicated, can be inferred in principle from the Bragg scattering of neutrons. Their elementary excitations, which are called magnons, are the normal modes of the coupled magnetic moments. They are bosonic quasiparticles and, possess gapless modes $-$ so-called Goldstone modes $-$ if the moments possess a continuous global symmetry. Suppose one found, in a neutron scattering experiment on a clean cubic magnet, an absence of Bragg peaks well below the Curie-Weiss temperature co-existing with linearly dispersing excitations while heat capacity measurements gave no indications of a phase transition. A remarkable possibility is that such a magnet might have no symmetry-broken order {\it at all}  and that the magnetic excitations above the ground state behave like charged particles interacting with linearly dispersing radiation. It is the purpose of this review is to explain how this possibility might be realized in pyrochlore magnets.

Such an unusual state of matter is one of many possible types of quantum spin liquid $-$ so named because quantum fluctuations are responsible for keeping the spins from entering a long-range ordered phase characterized by some broken symmetry even at zero temperature. Quantum spin liquids are to be contrasted with ordinary molecular liquids, which have only short-range order, and superfluid phases, which are symmetry broken phases. They are also to be contrasted with conventional paramagnets and also with collective paramagnets, or so-called classical spin liquids, which are finite temperature states exhibiting nontrivial short-range correlations. Quantum spin liquids are quite different: in common with fractional quantum Hall liquids, they appear disordered to local probes but they have some form of order which is instead ``encoded" nonlocally and which is often not characterizable in terms of symmetry. Not only are the natural characterising observables in quantum spin liquids nonlocal but they are often independent of anything but the topology of the nonlocal observable - such spin liquids are said to be ``topologically ordered".  For pragmatists, an essential feature of quantum spin liquids is that they are quantum phases exhibiting peculiar ``fractionalized" excitations meaning that the microscopic degrees of freedom are, for practical purposes, split into parts by the strong correlation or, more precisely, as a consequence of having long-range quantum entanglement in the ground state. \footnote{The precise definition of a quantum spin liquid is something that has drifted as people's understanding of them has grown. In their original incarnation, quantum spin liquids were magnets where zero point fluctuations about ordered states are of the order of the total spin moment so that the ground state is a superposition of states such that each spin has no net moment. Later, Anderson \cite{Anderson} proposed that liquid phases of paired spin singlets or valence bonds could form in frustrated magnets and, it was this, resonating valence bond (RVB) idea that set into motion the intense period of research into quantum spin liquids that continues to the present day.}

Until the mid-80's, condensed matter physicists had been able to understand a huge variety of different phases of matter $-$ indeed essentially all known states of matter $-$ in terms of symmetry and symmetry breaking. The discovery of the fractional quantum Hall effect in 1986 alerted condensed matter physicists to the importance of radically new concepts underlying the organisation of matter at low energies. Quantum spin liquids have been the main medium through which theorists have been able to generalise the physics of the fractional quantum Hall effect and the physics of low dimensional magnets and there has been immense progress in the understanding of these states of matter over the last thirty years. Yet no theoretical tool exists that will allow people to determine all possible ways in which matter can organise itself and for really ground-breaking insights we rely on guidance from experiment. It is for this reason that the experimental realisation of quantum spin liquids has been eagerly anticipated in the community  \cite{Balents_Nature}. One strategy to uncover these states of matter has been to explore magnets with magnetic ions sitting on lattices of corner-sharing triangles and tetrahedra. Antiferromagnetic exchange couplings between the ions are highly frustrated on these lattices such that any transition temperature occurs at a scale much lower than the Curie-Weiss temperature scale. In this way, one might hope that conventional long-range order is evaded entirely. A deeper reason for exploring geometrically frustrated magnets is that the semiclassical ground states are typically subject to a local constraint that can be interpreted as an emergent gauge invariance. Quantum fluctuations may then lead to a quantum spin liquid that is equivalent to a deconfined phase of a quantum mechanical gauge theory which, as we explain below, corresponds to a fractionalized phase.

At least the semiclassical part of this strategy is beautifully realized by a pair of materials which are becoming perhaps the archetypes of geometrical frustration among real magnetic materials: the spin ices Dy$_{2}$Ti$_{2}$O$_{7}$ and Ho$_{2}$Ti$_{2}$O$_{7}$ 
 \cite{SpinIceReview,BGH_Review,Gingras_Springer}. 
A great deal of recent theoretical and experimental effort has been devoted to exploring their rich behaviour at low temperatures where they enter a collective paramagnetic phase characterized by distinctive magnetic correlations that follow from a local constraint on the magnetic moments on each tetrahedron. 

The existence of spin ices is a promising state of affairs for the general research programme of evincing a quantum spin liquid in a magnetic material  \cite{Balents_Nature}. Whereas many proposed models with quantum spin liquid ground states are somewhat unphysical, as we explain below, quantum fluctuations acting on the set of spin ice states can be reasonably expected to lead to a quantum spin liquid with gapless photon-like excitations. We call this {\it quantum spin ice}. Furthermore, among the relatives of  spin ice materials, there are a number of materials where spin ice correlations exist at finite temperature and in which quantum fluctuations appear significant. The low temperature phases of these materials remain to be understood. 

This review is intended to bring together in one place an introduction to the quantum spin ice phase along with a survey of  those materials among the pyrochlore magnets most likely to harbour such a quantum spin liquid state. We begin, in Section~\ref{sec:spin_ice}, by reviewing some aspects of the physics of the classical spin ice because (i) the magnetism in these materials is the precursor state to quantum spin ice and (ii) a familiarity with the microscopic aspects of the pyrochlore magnets gained thereby will be invaluable in assessing the prospects for uncovering a quantum spin liquid among them. In Section~\ref{sec:QSI}, we review the arguments leading from an XXZ -like model on the pyrochlore lattice to an effective low energy description of the physics as an emergent electromagnetism. This proceeds in two mains steps $-$ by mapping from the spin model to a quantum dimer model (Section~\ref{sec:spin2loops}) and by mapping from the dimer model to a lattice gauge theory (Section~\ref{sec:loops2gauge}). We then describe the properties of the quantum spin ice phase (Section~\ref{sec:phenom}) and discuss both the stability (Section~\ref{sec:stability}) and naturalness (Section~\ref{sec:natural}) of this state of matter. We conclude by describing other contexts in which a U(1) liquid might be found in condensed matter systems (Section~\ref{sec:context}). The next main part of the review discusses quantum spin ice from a materials perspective. A discussion of the relevant microscopic features of candidate materials in Section~\ref{sec:materials_considerations} is followed by a description of the specific features of various candidate quantum spin ices including Tb$_{2}$Ti$_{2}$O$_{7}$ (\ref{sec:Tb2Ti2O7}), Pr$_{2}$M$_{2}$O$_{7}$ (M$=$Sn, Zr) (\ref{sec:Pr2M2O7}) and Yb$_{2}$Ti$_{2}$O$_{7}$ (\ref{sec:Yb2Ti2O7}).

\section{Spin Ice}

\subsection{Classical spin ice}

\label{sec:spin_ice}

\begin{figure}[h!]
\begin{center}
\includegraphics[width=7cm]{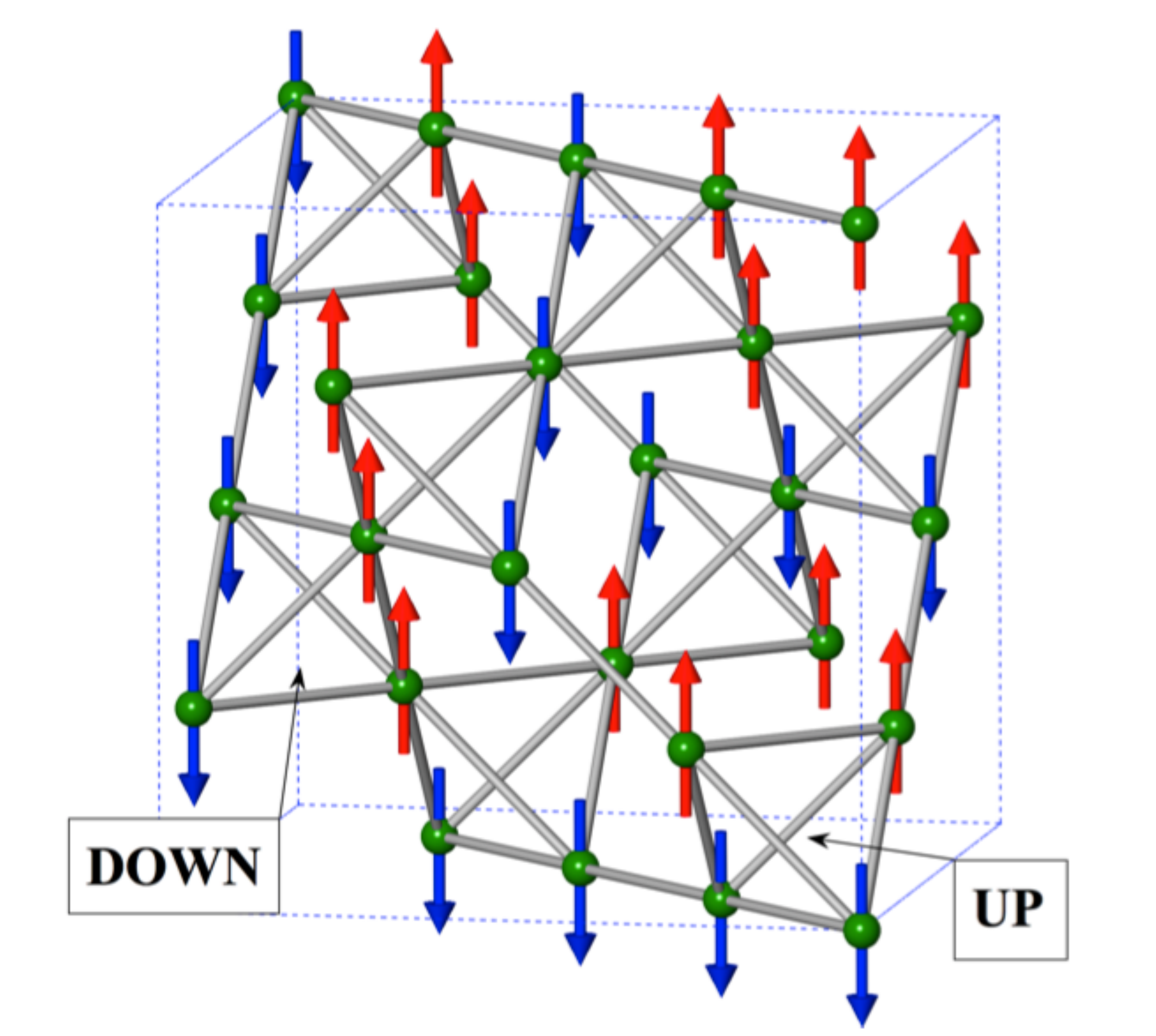}
\caption{The pyrochlore lattice with Ising magnetic moments on the lattice sites with a global $\hat{\mathbf{z}}$ axis anisotropy. The magnetic lattice can be  described as a face-centered cubic space lattice with either an ``up" or ``down" primitive basis. The figure shows a spin configuration fulfilling ``2-up"/``2-down" ice rules on each tetrahedron.
\label{fig:pyrochlore}}
\end{center}
\end{figure}

Our discussion of the physics of spin ice begins with the problem of classical Ising spins $S_i^z = \pm 1/2$ that reside on the sites of a pyrochlore lattice of corner-sharing tetrahedra (see. Fig.~\ref{fig:pyrochlore})
and interact among themselves via an antiferromagnetic nearest-neighbour exchange coupling $J_\parallel>0$. 
This model, first considered by Anderson in a 1956 paper \cite{Anderson_spinel}, was 
aimed at describing the magnetic ordering of spins on the octahedral sites of normal spinels and the related problem of ionic ordering in inverse spinels, both systems having a pyrochlore lattice.
The Ising antiferromagnet Hamiltonian, $H_{\rm I,AF}$, of Anderson's model is
\begin{equation}
\label{H_AF}
H_{\rm I,AF} = J_\parallel \sum_{\langle i,j \rangle} S_i^z S_j^z	,
\end{equation}
with $J_\parallel >0$ and where the sum is carried over the nearest-neighbour bonds of the pyrochlore lattice. Anderson found that $H_{\rm I,AF}$ admits an exponentially large number of ground states given by the simple rule that each ``up'' and ``down' tetrahedron (Fig.~\ref{fig:pyrochlore})
 must have a vanishing net spin. That is, on each tetrahedron, two spins must be ``up" 
and have $S^z = +1/2$ and two spins must be ``down" and have $S^z=-1/2$.
Anderson further recognized that this problem is closely related to that of hydrogen bonding in the common hexagonal (I$_{\rm h}$) phase of water ice or, more precisely, its cubic (I$_{\rm c}$) phase 
\footnote{Common water ice, I$_h$,  has a hexagonal structure, while the pyrochlore has cubic symmetry. The Ising pyrochlore 
problem is equivalent to cubic ice, I$_{\rm c}$, and not I$_{\rm h}$. This difference  does not modify the 
second ice-rule analogy or the connection between the statistical mechanics problem of proton coordination in water
ice and that of the spin arrangement in Anderson's model or spin ice below},
both being characterized by hydrogen (proton) configurations that obey the two Bernal-Fowler ice rules  \cite{Bernal_Fowler}.
The `second ice rule' is the one relevant to Anderson's model and to the main topic of this review
 \footnote{The first rule states that there should be only one proton per O$^{2-}$$-$O$^{2-}$ bond. 
This rule has no equivalent in Anderson's model and in spin ice discussed below \cite{SpinIceReview,BGH_Review}}.
The second rule states that for each fourfold coordinated oxygen O$^{2-}$ ion, there must two protons (H$^{+}$) near 
it and covalently bonded to that reference O$^{2-}$ ion, hence providing a hydrogen bond to two neighbouring O$^{2-}$ ions (hence H$_2$O molecules). 
At the same time, there are two protons far, which are themselves covalently bonded to the two other oxygen ions and hence hydrogen-bonded `back' to the original  reference O$^{2-}$ ion.
In other words, for each O$^{2-}$ ion, there are two protons near and covalently bonded to it and two farther protons hydrogen-bonded to it \cite{SpinIceReview,BGH_Review,Gingras_Springer}. This rule leads to an underconstrained system in regards to the number of minimum energy proton configurations in I$_{\rm h}$ and I$_{\rm c}$ and, in fact, leads to an exponentially large number of nearly degenerate proton configurations.
Linus Pauling had estimated in 1935  \cite{Pauling}
the number of ice-rule fulfilling ground states in water ice and the resulting low-temperature residual entropy $S_0$, 
finding close agreement with the experimental value being determined at about the same time by Giauque and co-workers  \cite{Giauque}. Pauling's reasoning for estimating $S_0$ can be adapted to Anderson's model in a number of closely related ways 
 \cite{SpinIceReview,BGH_Review,Gingras_Springer}. 
One finds $S_0 \sim (Nk_{\rm B}/2)\ln(3/2)$, where $N$ is the number of magnetic sites on the pyrochlore lattice and $k_{\rm B}$ is the Boltzmann constant.
Pauling's estimate is accurate within a few percent from the more precise estimate, both for I$_{\rm h}$  \cite{Nagle} and the Anderson model 
(or, equivalently, I$_{\rm c}$) \cite{Singh-Oitmaa}.

Anderson's antiferromagnetic (AF) model with Ising spins pointing along the $\pm \hat z$ direction (as in Fig.~\ref{fig:pyrochlore})
is unrealistic since there is no reason for the spins to prefer the $\hat z$ axis over the $\hat x$ or $\hat y$ axes in a system such as the pyrochlore lattice which has cubic symmetry. Consequently, that model did not much attract the attention of theorists and experimentalists investigating real magnetic materials for a long time. This changed with the 1997 discovery by Harris, Bramwell and co-workers of 
frustrated ferromagnetism in the insulating Ho$_2$Ti$_2$O$_7$ pyrochlore oxide  \cite{Harris_PRL,Harris_JMMM}.

\begin{figure}[h!]
\begin{center}
\includegraphics[width=7cm]{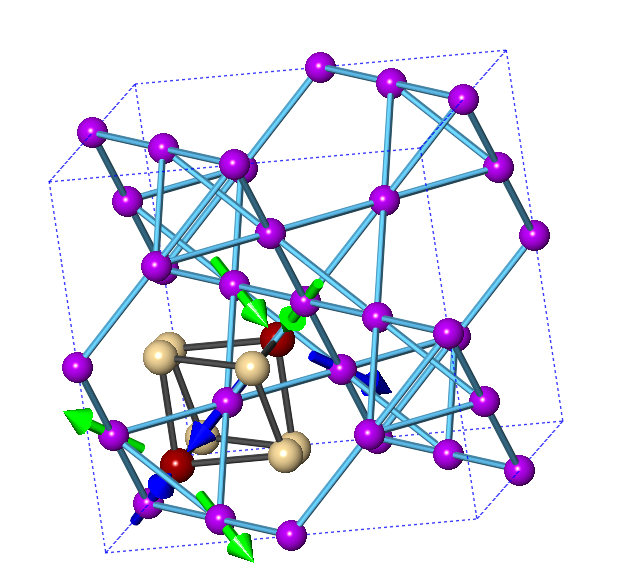}
\caption{The oxygen environment around a magnetic ion in R$_2$M$_2$O$_7$ materials with rare earth sites shown in purple. One distorted cube of oxygen ions is shown with the axial $[111]$ oxygens in red and the six transverse oxygens in beige. The six transverse oxygen ions are equidistant from the central rare earth site ($\sim 2.5$ \AA), lying further from the rare earth site than the axial oxygen ions which are at $\sim 2.2$ \AA.
\label{Ising_pyrochlore}}
\end{center}
\end{figure}

In Ho$_2$Ti$_2$O$_7$, the magnetic Ho$^{3+}$ ions reside on a pyrochlore lattice while Ti$^{4+}$ is non-magnetic. Despite overall ferromagnetic (FM) interactions characterized by a positive Curie-Weiss temperature, $\theta_{\rm CW} \approx +2$ K, Ho$_2$Ti$_2$O$_7$ was found to not develop conventional long-range magnetic order down to at least 50 mK  \cite{Harris_PRL,Harris_JMMM}.
 To rationalize this surprising behaviour, Harris and co-workers put forward the following argument.  Because of the strong local crystal electric field of trigonal symmetry acting at the magnetic sites, each Ho$^{3+}$ magnetic moment is forced to point strictly ``in''  or ``out'' of the two tetrahedra joined by that site, a direction that is along the pertinent cubic $[111]$ threefold symmetry axis that passes through the middle of the two 
site-joined tetrahedra (see Fig.~\ref{Ising_pyrochlore}). 
This large magnetic anisotropy allows one to describe the orientation of a Ho$^{3+}$ magnetic moment by an effective classical Ising spin $S_i^{z_i}$, with the $\hat z_i$ quantization direction now the local $[111]$ direction at site $i$. 
For a single tetrahedron with such a $[111]$ Ising spin at each of the four corners of a tetrahedron, and interacting with a nearest-neighbour FM coupling, the minimum energy is one of six states with two spins pointing ``in'' and two spins pointing ``out''. 
The ``up''/``down'' directions of Anderson's AF model in Eqn.~(\ref{H_AF}) 
now become the ``in''/``out'' directions of the frustrated FM Ising system  \cite{Harris_JPCM,Moessner_FM_PRB}. 
This point can be made clearer by considering the following simple model of interacting microscopic angular momenta ${\mathbf J}_i$
written as $H_{\rm FM}=-J_{\rm ex} \sum_{\langle i,j\rangle} {\mathbf J}_i \cdot {\mathbf J}_j$, where $J_{\rm ex}>0$  is ferromagnetic.
With ${\mathbf J}_i$ constrained to point along the local $\hat z_i$ [111] Ising direction, we have
${\mathbf J}_i = 2\langle {\rm J}^{z_i}\rangle S_i^z \hat z_i$
with $\langle {\rm J}^{z_i}\rangle$ the magnitude of the angular momentum  ${\mathbf J}_i$   
of the rare-earth ion.
Since $\hat z_i \cdot \hat z_j = -1/3$ on the pyrochlore lattice, we recover the model of Eqn.~(\ref{H_AF}) with
$J_\parallel = \frac{4J_{\rm ex}}{3} \langle {\rm J}^{z_i}\rangle^2$. 
The ferromagnetically coupled spins forced to point along local $\langle 111 \rangle$ directions thus become equivalent to the frustrated AF Ising model of Anderson in Eq.~(\ref{H_AF}).

However, unlike the AF model, the FM model with local $[111]$ Ising spins is physical since the crystal field responsible for the effective Ising nature of the magnetic moments is compatible with the cubic symmetry of the system (Fig.~\ref{Ising_pyrochlore}).
On the other hand, as in the AF model, the number of ground states in this frustrated Ising ferromagnet model 
 is exponentially large in the system size, resulting in an extensive residual entropy, again given approximately by the Pauling entropy $S_0$.
 In accord with this prediction, a number of magnetic specific heat measurements, and thus magnetic entropy, 
have confirmed that Ho$_2$Ti$_2$O$_7$  \cite{Cornelius} and Dy$_2$Ti$_2$O$_7$  \cite{Ramirez_Nature}, 
along with the Sn-variants, Ho$_2$Sn$_2$O$_7$  \cite{Ho2Sn2O7_entropy}
and Dy$_2$Sn$_2$O$_7$  \cite{Dy2Sn2O7_entropy} indeed possess a low-temperature residual entropy consistent with $S_0$.~\footnote{Recent efforts to explore the characteristic timescales from a.c. susceptibility within the correlated paramagnetic regime of Dy$_{2}$Ti$_{2}$O$_{7}$ have led to experiments measuring the heat capacity on similar time scales. These show the residual entropy falling significantly below the Pauling estimate \cite{Pomaranski}.}
Recent developments in high-pressure materials synthesis have allowed  people to make
Ho$_2$Ge$_2$O$_7$ and Dy$_2$Ge$_2$O$_7$, with these also displaying a residual entropy of $S_0$  \cite{R2Ge2O7_entropy}. Apart from holmium and dysprosium pyrochlores, the material Cd$_2$Er$_2$Se$_4$, spinel with magnetic erbium of pyrochlore sites, also exhibits a residual entropy $S_0$. \cite{Cd2Er2Se4}

Perhaps at least as interesting as the experimental determination of an $S_0$ residual entropy in these systems is the observation, 
originally made by Harris {\it et al.} \cite{Harris_PRL}, that the ``in''/``out'' Ising spins in the frustrated pyrochlore 
Ising ferromagnets can be seen to physically represent, or `map onto', 
the hydrogen bonding or, more precisely, the proton displacement with respect
to the midpoint between two O$^{2-}$ ions in water ice  \cite{SpinIceReview,BGH_Review,Gingras_Springer}. 
This observation led Harris, Bramwell and co-workers to coin the name {\it spin ice} for these systems.

While the frustrated nearest-neighbour FM Ising model helped rationalize the thermodynamic and magnetic properties of spin ice compounds
  \cite{Harris_PRL}, it was nevertheless initially found rather puzzling why these materials should be described by such a simple model or even possess a residual low-temperature magnetic entropy equal to $S_0$  \cite{denHertog}. 
While the resolution of this paradox for classical spin ice  may at first hand appear peripheral to the topic of this review, 
it will prove of significant importance and rich in physical insight when we later discuss the low-energy excitations in quantum spin ice.

In the Ho and Dy based spin ice compounds, the magnetic moment ${\mathbf \mu}$ of the Ho$^{3+}$ and Dy$^{3+}$ ions 
is of the order of $\mu \sim 10$ $\mu_{\rm B}$ \cite{Siddharthan}. Considering a nearest-neighbour distance $r_{\rm nn} \sim 3.6$ $\AA$ in (Ho,Dy)$_2$(Ti,Sn,Ge)$_2$O$_7$ spin ices, one finds the scale for the magnetostatic dipolar interactions among nearest neighbours to be $D\sim 1.4$ K \cite{denHertog,Siddharthan}. 
We thus have $D\sim \vert \theta_{\rm CW} \vert$ in these systems and dipolar interactions must therefore
 be considered carefully at the outset when discussing spin ice physics in real materials  \cite{denHertog}. Crucially, magnetostatic dipole-dipole interactions display two key features that would seem to make them antagonistic to the formation of a degenerate low-temperature state with extensive entropy. Firstly, they are very long-ranged, decaying as
 $1/{|{\mathbf r}_{ij}|^3}$
with the separation distance $\vert {\mathbf r} _{ij} \vert$ between magnetic moments. 
Secondly, they strongly couple the direction of the magnetic moments, ${\bm \mu}_i$ and ${\bm \mu}_j$, at position ${\mathbf r}_i$ and ${\mathbf r}_j$, respectively, 
with the relative position vector ${\mathbf r}_{ij} \equiv {\mathbf r}_j  - {\mathbf r}_i$.
Both properties would naively appear to dramatically 
constrain admissible minimum energy orientations of the moments well beyond the ice rule imposed by the nearest-neighbour model of Eqn.~(\ref{H_AF}).
Interestingly, one finds that dipolar interactions between local 
$\langle 111 \rangle$ Ising moments truncated at the nearest-neighbour distance are ferromagnetic-like, as in the Harris {\it et al.} frustrated Ising model  \cite{Harris_PRL,Harris_JPCM,Moessner_FM_PRB}.
 This dipolar spin ice model (DSIM) \cite{denHertog}, and its subsequent generalization, 
including additional short-range Ising exchange interactions \cite{Ruff_PRL,Yavorskii},
has proven highly successful at explaining quantitatively a wide variety of behaviours displayed by spin ice compounds.
Perhaps most noteworthy about the DSIM is the observation, first made through 
Monte Carlo simulations  \cite{denHertog}, that the long-range part of the dipolar interactions beyond the nearest neighbour distance 
appears ``self-screened'' \cite{Gingras_CJP}. That is, they only lead to a transition to long-range order at a temperature 
$T_c \sim 0.07 D/k_{\rm B}$ \cite{Melko2001,Melko2004}.
This critical temperature  is  much lower than the temperature $T_{\rm SI} \sim D/k_{\rm B}$ 
at which the system crosses over from the trivial paramagnetic state to the spin ice regime, characterized by a fulfilment of the ice rules,
and marked by a broad peak in the magnetic specific heat  \cite{Ramirez_Nature,denHertog}.~\footnote{This argument omits the nearest neighbour exchange interaction $J$. For positive ferromagnetic $J$ the crossover temperature $T_{\rm SI}\sim J/3+5D/3$  \cite{SpinIceReview,BGH_Review,Gingras_Springer,denHertog}.}
This behaviour originates from a remarkable feature of the anisotropic nature of the dipolar interactions on the pyrochlore lattice \cite{denHertog,Gingras_CJP}. 
Namely, on this lattice, magnetostatic dipole-dipole interactions differ very little, and only at short distance, from a model Hamiltonian with nearest neighbour interactions with the same ground state degeneracy as that prescribed by the ice rules  \cite{Isakov_PRL}.  
Of the two formulations explaining that phenomenology  \cite{Isakov_PRL,Castelnovo2008,Castelnovo_AnnRev}, 
the one that we discuss next is the most relevant to the topic of this review.

Perhaps the formulation that has been most simply able to capture in one sweep the key 
features of the DSIM is the so-called dumbbell model of Castelnovo, Moessner and Sondhi  \cite{Castelnovo2008}.
In essence, the dumbbell model is a way of visualizing a multipole expansion about diamond lattice sites. It takes for a start the point-like magnetic dipoles of the DSIM and fattens them up into a rod with two magnetic charges $\pm q_m$ centered on the two tetrahedra connected by a $[111]$ Ising spin.
The centers of the tetrahedra form a diamond lattice with lattice spacing  $a_d$. The magnetic charge $q_m$ is chosen so that, given 
the diamond center-to-center distance $a_d$, one recovers the original magnetic dipole moment with $\mu \equiv q_m a_d$. 
The original model of point-like dipoles has now been recast as a model of magnetic charges interacting though a ``magnetic Coulomb potential'', 
the latter model having, by the construction $\mu \equiv q_m a_d$, the same $1/r^3$ dipolar far-field as the DSIM.
Since in this model the magnetic charges, or ``monopoles'', from the four dumbbells at the center of 
a given tetrahedron overlap, 
a regularization of the Coulomb potential is introduced  \cite{Castelnovo2008}. 
The strength of the regularization potential for overlapping monopoles is adjusted so as to
correctly  recover the interaction energy for nearest-neighbour dipoles along with the contribution coming from a nearest-neighbour exchange 
 \cite{Castelnovo2008,Castelnovo_AnnRev}.

As a result of these constructions, the dumbbell Hamiltonian $H_{\rm db}$ can thus be expressed in 
 terms of the net charge $Q_\alpha = \sum_{i\in \alpha}  = 0, \pm 2q_m$ and $\pm 4q_m$ at the center of the $\alpha$'th tetrahedron, with 
\begin{equation}
\label{eq:Coulomb}
H_{\rm db} = \frac{\mu_0}{4\pi} \sum_{\alpha > \beta} \frac{Q_\alpha Q_\beta}{r_{\alpha\beta}} + \frac{v_0}{2} \sum_\alpha {Q_\alpha}^2,
\end{equation}
where the onsite term $v_0$ is determined from the condition that this model should reproduce the spin flip energy of the dipolar spin ice model. 
From this formulation, it is clear that
in the limit $v_0 \rightarrow \infty$,  the ground states of the system have $Q_\alpha=0$ for all $\alpha$.
These are precisely the two-in/two-out ice-rule obeying states. 
For finite $v_0$, the low-temperature state of the system
is ultimately determined by the competition between the self-energy cost 
of a charge and the energy gain of a specific arrangement of positive and
negative magnetic charges on the diamond lattice.
This dumbbell model describes rather accurately the DSIM, in particular the long-range $1/r^3$ nature of its dipolar part.
 Most importantly, the dumbbell model has the same  quasi-degenerate ice-rule obeying states as the nearest-neighbour frustrated 
Ising model of Harris {\it et al.} and displays, in particular, a residual
low-temperature Pauling entropy $S_0$.

Noting that the condition of vanishing charge $Q_\alpha=0$ is equivalent to the ``2-in''/``2-out'' spin configuration on each tetrahedron, we may introduce a coarse-grained ``spin field" which we denote as ${\mathbf B}$ for a reason that will become clear in the next section. Then the statement that $Q_\alpha=0$ is akin to stating that the ``spin field" has zero divergence, $\nabla\cdot\mathbf{B}=0$. When there is no local source or sink of the coarse-grained ``spin field'' in classical spin ice, the magnetic structure
factor, which can be measured experimentally via neutron scattering, exhibit singularities 
in reciprocal space at nuclear Bragg points  \cite{Henley2005,Henley2010}.
The intensity profile around these singularities has a characteristic ``bow-tie" form in planes through the singular points which have come to be known as ``pinch-points'' \cite{Henley2010,Fennell_Science}.

For a reason that will be expanded upon in Section 3, one refers to the  $Q_\alpha \neq 0$ charges,
sources and sinks of the spin field $\mathbf{B}$,  as gauge charges.
The beauty of the dipolar spin ice is that, thanks to the underlying dipolar interactions, the gauge charge is also a magnetic charge, or ``monopole''
that is a sink or source of the local magnetization field, \cite{Castelnovo2008}, since each spin comes along with its magnetic moment $\mu$.
So one perspective on the effective low-energy theory of dipolar spin ice 
is that of a magnetic Coulomb gas, with charges on a diamond lattice and 
in the grand canonical ensemble  \cite{Jaubert_NatPhys,Jaubert_JPCM}.
The flipping of an individual Ising spin thus 
corresponds to the nucleation of two magnetic charges of opposite sign which 
can then thermally diffuse while interacting with an effective emergent Coulomb potential.
In short, at low-temperature and at low-energy, the long-distance physics of dipolar spin ice
 is equivalent to a magnetic formulation of Gauss's law in which 
the elementary excitations interact with  $1/r_{\alpha\beta}$ ``magnetic" Coulomb potential as given by Eqn.~(\ref{eq:Coulomb}). 
As we shall see in Section 3, the present magnetic
 charges correspond to the classical limit of the magnetic charge of the $U(1)$ 
description of the quantum spin liquid. In the present case of the classical dipolar spin ice, these are a
remarkable manifestation of the ``fractionalization'' of the individual elementary dipole moment (spin) 
flip excitation and, because of the
$1/r_{\alpha\beta}$ nature of their mutual Coulomb interaction, the work to separate  two charges of opposite sign
 is finite and the charges are said to be ``deconfined''  \cite{Castelnovo2008,Castelnovo_AnnRev}.

The paper \cite{Castelnovo2008}, highlighting the existence of monopole excitations of energetic origin in the dipolar spin ice materials, has led to a number of experiments aimed at probing their effects on various magnetic properties. Because this chapter in the history of spin ices fits neatly into a discussion of the experimental probes of exotic excitations, we briefly describe a cross-section of this work.
We direct the reader to recent reviews for a more extensive discussion  \cite{Castelnovo_AnnRev,Henley2010}.
For the purpose of this review, it suffices to say that, 
while the description of the low-energy excitations in dipolar spin ice model in  terms of monopoles 
is most likely correct, 
few experiments can be said to have positively exposed monopoles in classical spin ice compounds.
Some first generation experiments using muon spin relaxation  \cite{Bramwell_Nature}
and relaxation of the bulk magnetization  \cite{Giblin_NatPhys} 
have later been subject to a controversy in the former case  \cite{Dunsiger_PRL,Blundell_PRL,Sala_PRL} 
and the interpretation critiqued in the latter  \cite{Revell_NatPhys}. 
The temperature dependence of the relaxation time extracted from magnetic AC susceptibility 
has been partially rationalized in terms of diffusive motion of
monopoles  \cite{Jaubert_NatPhys,Jaubert_JPCM}, 
but such a description becomes less compelling deep in the spin ice regime  \cite{Revell_NatPhys}, 
either because of complexities arising from sample quality and/or disorder issues \cite{Revell_NatPhys}
or yet other aspects of the real materials that have not been incorporated into the magnetic Coulomb gas theoretical framework 
 \cite{Jaubert_NatPhys,Jaubert_JPCM}.
From another perspective, we note that the assignment of the temperature dependence of the width 
of the neutron scattering lineshape near the so-called ``pinch points'' in the Ho$_2$Ti$_2$O$_7$ material  \cite{Fennell_Science}
to the thermal nucleation of monopoles with separations of $O(10^2)$ $\AA$  
is not evident. The reason for that is that the lowest temperature considered in Ref. [\onlinecite{Fennell_Science}]
was barely below $T_{\rm SI}\sim 1.9$ K at which this material 
enters the spin ice state  \cite{Bramwell2001}.
 \footnote{The existence of a pinch point at $T\gg T_{\rm SI}$,  deep in the paramagnetic regime, is presumably due to the pinch-point singularity arising directly from the anisotropic $1/r^3$ dipolar interaction \cite{Sen}.} 
The monopole-based description of the neutron scattering data of 
Dy$_2$Ti$_2$O$_7$ in a magnetic field along the $[100]$ direction is, at this time, 
perhaps the one that most compellingly endorses the picture of fractionalized monopoles in these materials  \cite{Morris_Science}. While not direct evidence for their existence, the field-driven first order metamagnetic transition for a field along the $[111]$ direction is nicely and simply rationalised in terms of a crystallisation of positive and negative monopoles \cite{Castelnovo2008,Kadowaki}.

In view of the prospect of ultimately identifying quantum spin ice candidate materials 
characterized by another gapped and fractionalized excitation, in addition to the magnetic monopole, as well as an emerging gauge boson (``photon''), 
 new experimental methodologies with unambiguous signatures for these various excitations need to be developed. 
Such techniques could be first benchmarked on 
classical dipolar spin ices and perhaps proved successful in achieving an 
explicit demonstration of the existence of the aforementioned ``monopoles''.
Having reviewed the topic of classical spin ices, with a focus on the description 
of their low temperature behaviour in terms of fractionalized monopoles,
we now move onto the heart of this review -- the topic of quantum spin ice.

\subsection{Naming Conventions}
\label{sec:conventions}

The reader who wishes to study the original literature on classical and quantum spin ice will come across a number of different naming conventions for the emergent particles and fields. We give here a quick guide to the main conventions and, at the same time, fix our own.
As stated above, in classical spin ice, the ice constraint on each tetrahedron can be thought of as a divergence free condition on a coarse-grained field. In the literature one can find the various naming conventions given in the top panel of Table~\ref{tab:names}. Point-like defects resulting from the local breaking of the ice constraint are sources of the coarse-grained field. Since the emergent charges have a physical magnetic response, we call them magnetic charges or monopoles in common with the classical spin ice literature  \cite{Castelnovo2008,Castelnovo_AnnRev}. 

A second kind of gapped excitation appears in quantum spin ice which we call a vison. The nature of this excitation can be seen most clearly from the compact gauge theory discussed in Section~\ref{sec:loops2gauge}. In this effective field theory \cite{Hermele2004}, the spin ice constraint appears as Gauss' law in an electric field defined on links of the diamond lattice. The effective field theory has both electric and magnetic degrees of freedom. In the convention of Hamiltonian lattice gauge theory, the vison excitation is a source of magnetic flux and is typically called a magnetic monopole in that community. At this point, confusion might easily arise. We have tried to keep our notation consistent with that, so far, predominantly employed by classical spin ice researchers and, for this reason, have swapped the convention $-$ the vison appears as a source of electric field while the magnetic monopole keeps its identity as the emergent charge familiar from classical spin ice \cite{Castelnovo2008}. We keep ourselves, however, from calling the vison an electric charge because, while the magnetic monopole is indeed a source of the physical magnetic field (or, more correctly the macroscopic field $\mathbf{H}$),  the vison is a merely a source of the fictitious electric field in the lattice gauge theory.

\onecolumngrid
\begin{table*}[htpb]
\centering
\begin{tabular}{cl}
\hline
\hline
\multicolumn{2}{c}{{\bf Coarse-grained field present also in classical limit}} \\
\hline\hline
\textcolor{blue}{Magnetic field} $\mathbf{B}$ or $\mathbf{H}$ & [\onlinecite{Castelnovo2008}] \\
Polarization $\mathbf{P}$  & [\onlinecite{Henley2005}] \\
Spin field $\mathbf{S}$ & [\onlinecite{Isakov_PRL}] \\
Electric field $\mathbf{e}$ & [\onlinecite{Hermele2004}] \\
\hline\hline
\multicolumn{2}{c}{{\bf Flipped spin defects/ Emergent charges on diamond lattice}} \\
\hline\hline
\textcolor{blue}{Magnetic monopole/charge} & [\onlinecite{Castelnovo2008}] + spin ice literature since 2008 \\
Electric charge   &  [\onlinecite{Hermele2004}]  from gauge theory literature \\
Spinon  & [\onlinecite{Savary2011}] from quantum spin liquid literature  \\
\hline\hline
\multicolumn{2}{c}{\bf {Gapped topological defects} } \\
\hline\hline
\textcolor{blue}{Vison} & [\onlinecite{Moessner2003}] and quantum spin liquid literature \\
Magnetic monopole  & [\onlinecite{Hermele2004}] from gauge theory literature \\
\hline\hline
\end{tabular}
\caption{Naming conventions for excitations and fields in quantum spin ice. Our conventions are highlighted.}
\label{tab:names}
\end{table*}
\twocolumngrid

The vison borrows its name from the literature on quantum spin liquids where, to our knowledge, it first appeared to describe fluxes in $Z_2$ gauge theory \cite{Senthil}. This literature would also naturally assign the term spinon (or fractionalized spin) to the charges that coherently hop on the diamond lattice sites and which decohere into the magnetic monopoles of classical spin ice at finite temperature.

\begin{figure}[h!]
\begin{center}
\includegraphics[width=7cm,angle=-90]{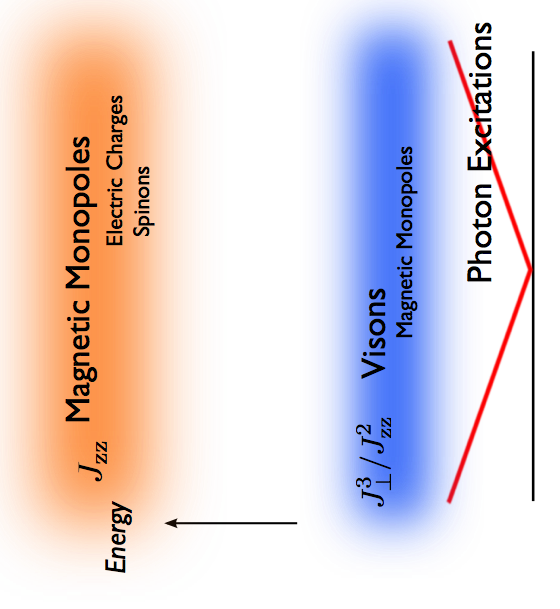}
\caption{Schematic of the spectrum of excitations in quantum spin ice including the approximate energy scales and different naming conventions.  \label{fig:Excitations}}
\end{center}
\end{figure}

Figure~\ref{fig:Excitations} shows how the excitations are organised by energy scale in quantum spin ice.

\section{Quantum Spin Ice}
\label{sec:QSI}

Quantum spin ice is a type of $U(1)$ quantum spin liquid which might be observed in certain pyrochlore magnets. A $U(1)$ spin liquid in three dimensions is a collective paramagnetic phase of matter with fractionalized excitations at low energy that are {\it gapless}, with linear dispersion $\omega\sim \vert k\vert$ and with two transverse polarizations. In short, these excitations behave like particles of light. From the standpoint of modern relativistic quantum field theory, physicists regard gauge invariance, and hence electromagnetic radiation, as being the inevitable consequence of having a quantum theory of relativistic massless spin one particles \cite{Weinberg_I}. In the present condensed matter context, the reasoning is turned around: the quantum spin liquid has an emergent low energy gauge redundancy so that localised magnetic excitations of spin one, which have no preferred axis because there is no spontaneous 
symmetry breaking, lose one polarization and behave instead like particles of light. In contrast to magnons in long-range magnetically ordered phases, which have two polarisations because one direction is fixed by the broken symmetry, the fact of having two polarizations of photon excitations, whether fundamental or emergent, is enforced by gauge invariance.  In this section, we shall see in a little more detail how magnetic interactions may give rise to a variant of ordinary quantum electrodynamics. We then describe various properties of these exotic phases and review some of the ways in which they might be probed experimentally in real quantum magnets. Next, we consider the naturalness of quantum spin ice models and discuss from a very general perspective, the prospects of seeing quantum liquids of this type in real materials.
 We conclude with a short section (Section~\ref{sec:context}) mentioning other possible condensed matter realizations of $U(1)$ 
liquids as well as putting quantum spin ice into the broader context of understanding quantum spin liquid phases.

\subsection{From a spin model to loops}
\label{sec:spin2loops}

\begin{figure}[h!]
\begin{center}
\includegraphics[width=7cm]{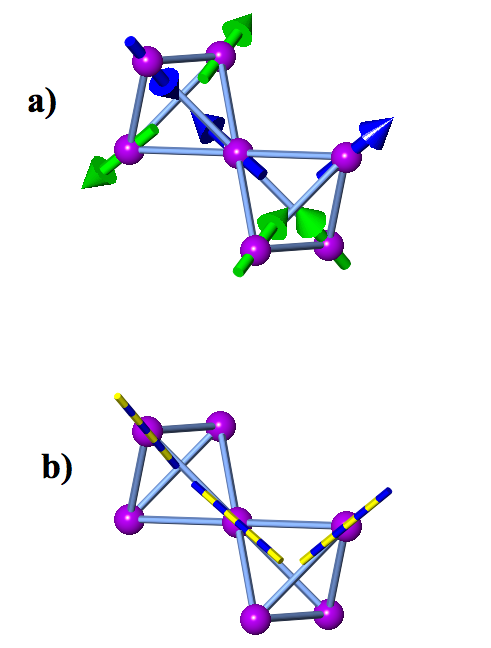}
\caption{Figure showing how the loop manifold for the dimer model is constructed from the spin ice states.  The figure at the top shows one particular spin ice configuration on a pair of tetrahedra on the pyrochlore lattice and the lower figure shows the corresponding dimer
links of the diamond lattice connecting the centres of the pyrochlore tetrahedra. The spin configuration at the top is mapped to the dimer configuration (blue/yellow rods) in the lower figure as follows. The diamond lattice is bipartite so that alternating pyrochlore tetrahedra can be labelled A and B. Suppose the left tetrahedron is an A tetrahedron. The rule to make a loop configuration is to lay a dimer along a diamond lattice link when a moment points into an A tetrahedron (or out of a B tetrahedron). Evidently, the ice states lead to dimer configurations where exactly two dimers meet at each diamond lattice site. The set of such states are acted on by the quantum dimer Hamiltonian $H=H_{\rm Dimer} + H_{\rm N}$ given in the main text. \label{fig:dimer}}
\end{center}
\end{figure}

We begin by returning to classical spin ice because it is, in some sense to be made more precise later, the precursor state to the quantum spin liquid state of quantum spin ice systems. Also, it will give us a classical example in which a $U(1)$ gauge redundancy appears, or really emerges,
 at low energies in a magnet. The key to making a spin ice is to frustrate an Ising model by putting it on a pyrochlore lattice (see Fig.~\ref{Ising_pyrochlore}). 
As discussed earlier in Section \ref{sec:spin_ice}, in real magnets, the Ising spins interact in spin space as though they were pointing along the local 
$\langle 111\rangle$ directions. The interactions in a classical nearest-neighbour spin ice (CSI) model are described by the 
same Hamiltonian as in Eqn.~(\ref{H_AF}) that we rewrite here:
\begin{equation}  H_{\rm CSI} =  J_{\parallel} \sum_{\langle ij\rangle} S^{z}_{i}S^{z}_{j}. 
 \label{eq:SI}  
\end{equation}
To emphasize something we have already mentioned: this classical Hamiltonian has a hugely degenerate ground state composed of spin configurations fulfilling the ``ice rule" of two spins pointing in and two pointing out of each tetrahedron as illustrated in 
Fig.~\ref{Ising_pyrochlore} and the top panel of Fig.~\ref{fig:dimer}. We denote the Hilbert space of ice states as $\mathcal{I}$. The spectrum of states has a gap of $4J_{\parallel}$ to flipped spin defects. The ice rule can be formulated as $\sum_{a} {\bm S}_{a}\cdot\mathbf{\hat{z}}_{a}=0$ (where the sum runs over all the sublattice sites $a$ of a tetrahedron) for each tetrahedral element of the pyrochlore lattice. This condition is a zero divergence condition on a lattice \cite{Henley2010}
which may be coarse-grained to $\nabla\cdot\mathbf{B}=0$ where the ``magnetic field" $\mathbf{B}$ is a coarse-grained analogue of the spin field ${\bm S}_{a}$ on the lattice. Since any vector field can be decomposed into the sum of two fields with, respectively, zero divergence and zero curl, in order to obtain thermodynamic quantities within the restricted ($\nabla\cdot\mathbf{B}=0$) manifold of spin ice states, one must average solely over the circulation of $\mathbf{B}$. In dramatic contrast to conventional long-range ordered magnets at low energies, in spin ices, this coarse-grained circulation is unconstrained and runs over a number of states that scales as $\exp(\alpha V)$ in the volume $V$ of the system. We can look at the divergence-free constraint as an emergent gauge invariance, since one may introduce a vector potential $\mathbf{A}$ such that $\mathbf{B}=\nabla\times\mathbf{A}$ and could carry out gauge transformations on $\mathbf{A}$ that would leave the divergence-free condition invariant. The divergence-free condition is thus a characteristic of the ground states of  Eqn.~(\ref{eq:SI}). At finite but low temperatures, this condition is weakly violated
 by the thermal excitation of spin flip defects (i.e. the ``monopoles'' of the classical dipolar spin ice). 
As the electrostatic analogy suggests, these defects behave like sources of $\mathbf{B}$ and, at temperatures where such effective charges are dilute, spin ice should behave much like a dilute plasma described in the grand canonical ensemble. This physics becomes richer still when the underlying microscopic magnetic moments, $\bm{\mu}_{a}\propto {\bm S}_{a}$, interact through a long-range dipolar coupling $-$ hence the review of dipolar spin ices in Section~\ref{sec:spin_ice}. In particular, 
as discussed in Section~\ref{sec:spin_ice},
the dipolar interaction about the spin ice background fractionalizes into an energetic Coulomb interaction between defects in a background of tetrahedra satisfying the spin ice rule  \cite{Castelnovo2008}.

\begin{figure}[h!]
\begin{center}
\includegraphics[width=7cm]{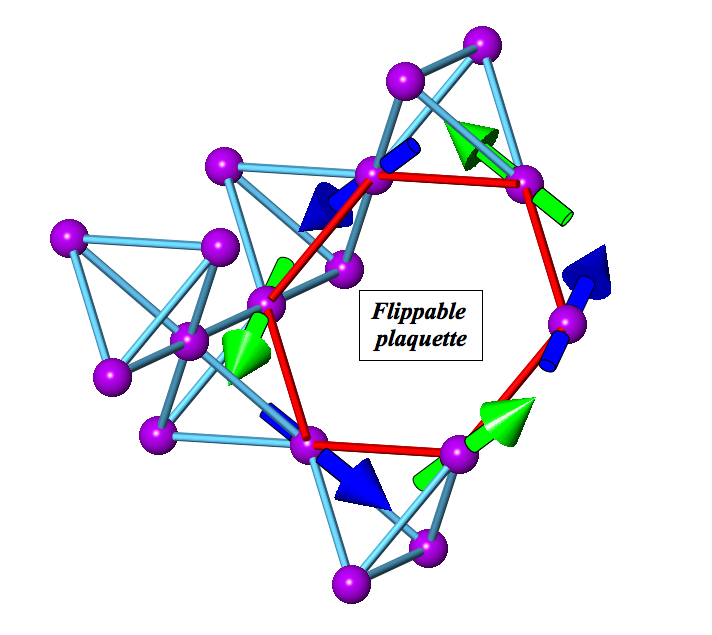}
\caption{Closed hexagonal loop on the pyrochlore lattice and on the diamond lattice. The figure shows a segment of a $[111]$ kagome plane in the pyrochlore lattice showing six tetrahedra with the diamond lattice sites (at the centres of the tetrahedra) and the hexagonal loop connecting the diamond sites (in red). \label{fig:hexagon}}
\end{center}
\end{figure}

The gauge invariance of classical spin ice turns out to be crucial to the quantum case to which we now turn. We now allow for the presence of (perturbative) ``transverse''  nearest-neighbour exchange couplings in addition to the ``longitudinal'' (Ising) 
exchange part defined by Eqn.~(\ref{eq:SI}). Our only requirement is that the transverse couplings should have a characteristic energy scale $J_{\perp}\ll J_{\parallel}$ so that there is little mixing of the ice rule states with 
canted spin states away from the local $[111]$ Ising direction. 
We shall discuss in Section~\ref{sec:materials_considerations} the most general nearest-neighbour  
symmetry-allowed  anisotropic Hamiltonian on the pyrochlore lattice that does not commute with $H_{\rm CSI}$ and  thus causes quantum dynamics.
For now, we consider a minimal spin model that contains 
quantum dynamics within a spin ice state and which is a sort of local XXZ model with transverse coupling $J_\perp$.
\begin{equation}
  H_{\rm QSI} \equiv H_{\rm CSI} + H_{\perp} = H_{\rm CSI} - J_{\perp} \sum_{\langle ij\rangle} (S_i^+ S_j^- + S_i^- S_j^+)
\label{eq:QSI}  
\end{equation}
We shall comment, in Section~\ref{sec:natural}, on the conditions under which real materials may 
exhibit such a $J_\perp \ll  J_\parallel$ separation of energy scales. 

The reason for considering such a separation of scales 
is that we require the manifold of classical spin ice states $\mathcal{I}$,
which form a reference classical spin liquid  \cite{Balents_Nature}, 
to be the background on which quantum fluctuations can act perturbatively. 
This allows one to carry out perturbation theory in the transverse couplings $-$ the zeroth order states being the whole manifold $\mathcal{I}$ of degenerate ice states. 
The lowest order terms derived from a canonical perturbation theory that preserve the ice rule constraint are ring exchange terms that live on the hexagonal loops on the pyrochlore lattice (Fig.~\ref{fig:hexagon}). 
Up to a numerical prefactor, the effective low energy Hamiltonian that describes quantum fluctuations within $\mathcal{I}$ is
\begin{equation}  
\label{eq:ringexchange}
H_{\rm ring} \sim \frac{J_{\perp}^{3}}{J_{\parallel}^{2}} 
\sum_{h \in \left\{\hexagon\right\} } S^{+}_{h,1} S^{-}_{h,2}  S^{+}_{h,3}  S^{-}_{h,4}  S^{+}_{h,5}  S^{-}_{h,6}  + {\rm h.c.}  ,
\end{equation}
where the sum is taken over the set of all hexagonal plaquettes in the pyrochlore lattice labelled by $\left\{\hexagon\right\}$. It turns out that the sign of the ring exchange coupling $\mathcal{J}_{\rm ring}\sim J_{\perp}^{3}/J_{\parallel}^{2}$ is not important since one can unitarily transform one sign to the other \cite{Hermele2004} \footnote[1]{The transformation that changes the sign of the ring exchange changes also the background flux through the hexagonal plaquettes. This does not change the low energy physics. However, in the gMFT \cite{Savary2011,Lee2012} described later in this article, the background $\pi$ flux affects the dispersion of spinon excitations which, in turn, can affect the phase diagram in the presence of competing interactions.}. 
The $U(1)$ liquid phase of the quantum spin ice model arises from this effective Hamiltonian. The ring exchange term has a local $U(1)$ gauge invariance: one can perform a rotation of the spin coordinate frame about the local $[111]$ $\hat z$ axis on a given tetrahedron $\prod_{i\in t} \exp(i\theta S^{z}_{i})$, where the product is taken over sites $i$ on pyrochlore tetrahedron $t$, which may capture $0$ or $2$ vertices of a hexagonal ring exchange operator. This rotation leaves the effective ring exchange Hamiltonian of Eq.~(\ref{eq:ringexchange}) unchanged. The notation $U(1)$ refers to the fact that the local transformation that leaves the Hamiltonian invariant is an element of the $U(1)$ group of $1\times 1$ unitary matrices (complex phases). More generally, to all orders in the perturbation expansion in $J_{\perp}$, there is a local gauge invariance of this sort.


\begin{figure}[h!]
\begin{center}
\includegraphics[width=9cm]{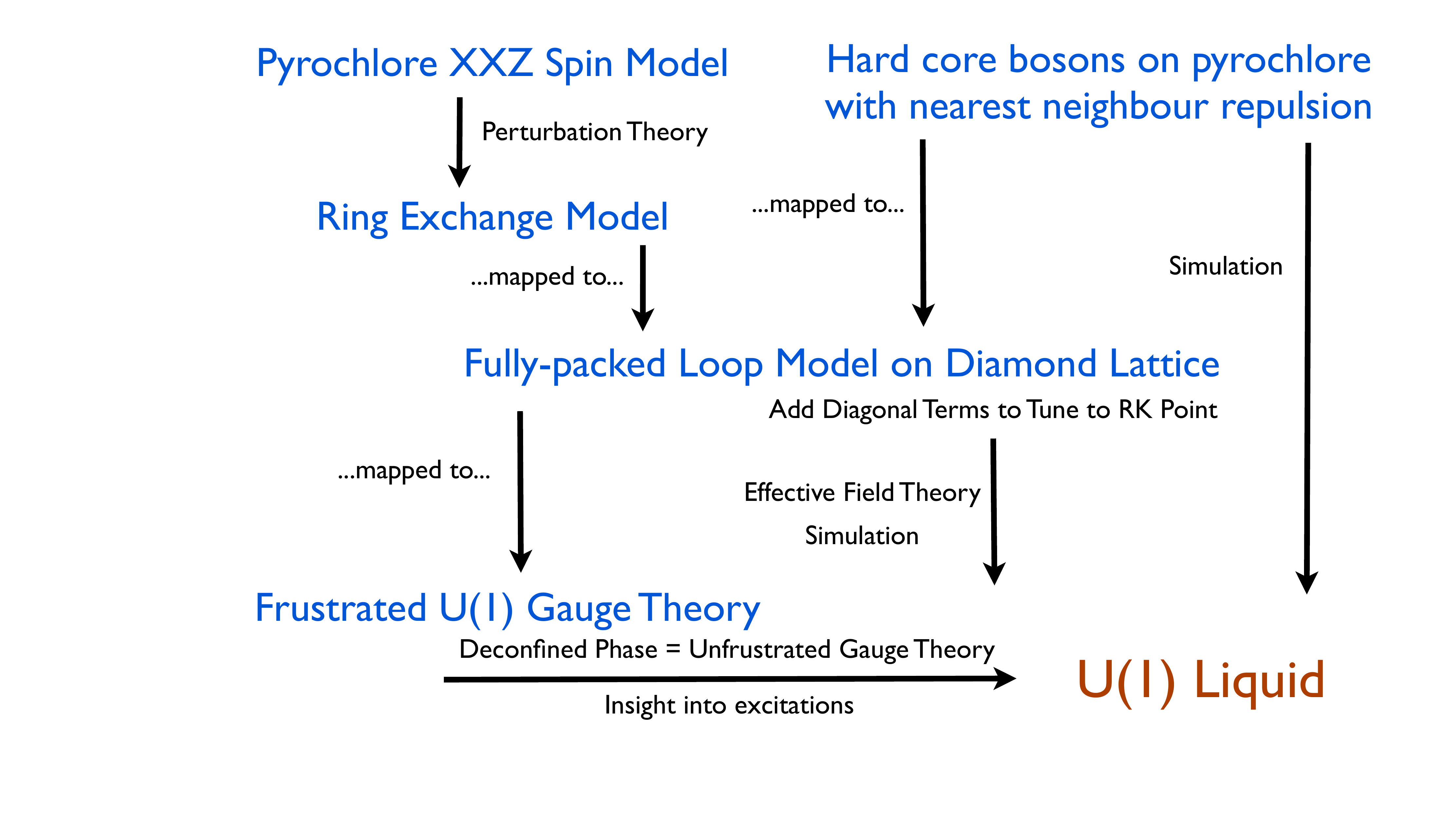}
\caption{Figure showing the conceptual relationships between various models mentioned in the review and which have been instrumental to understanding the $U(1)$ liquid phase in pyrochlore lattice magnets. \label{fig:analysis}}
\end{center}
\end{figure}

Now that we have motivated the existence of dominant ring exchange terms in certain effective low energy Hamiltonian models of pyrochlore magnets, we seek to understand the resulting low energy phase.
This discussion will take a while and involve ramified ideas.
 The problem of understanding the ring exchange model has been tackled in several different ways that are  summarized in Figure \ref{fig:analysis}. In particular, it is helpful to recognize the ring exchange model as a type of quantum dimer model on the diamond lattice as illustrated in Fig.~\ref{fig:dimer}. We refer the reader to Refs. [\onlinecite{Hermele2004, Shannon2011,Benton2012}] and to Fig.~\ref{fig:dimer} for further details of the mapping of the dimer model. The main idea is that the spin ice states correspond to so-called ``loop coverings'' of the diamond lattice where dimers on the links of the diamond lattice are placed end-to-end. In this representation, the spin ice rule is equivalent to the constraint that every diamond site has exactly two dimers connected to it. The pyrochlore ring exchange Hamiltonian of Eq.~(\ref{eq:ringexchange}) can be written in a form that symbolizes the quantum dynamics of these dimers
\begin{equation}
\label{eqn:dimer_model}
\includegraphics*[width=0.6\linewidth]{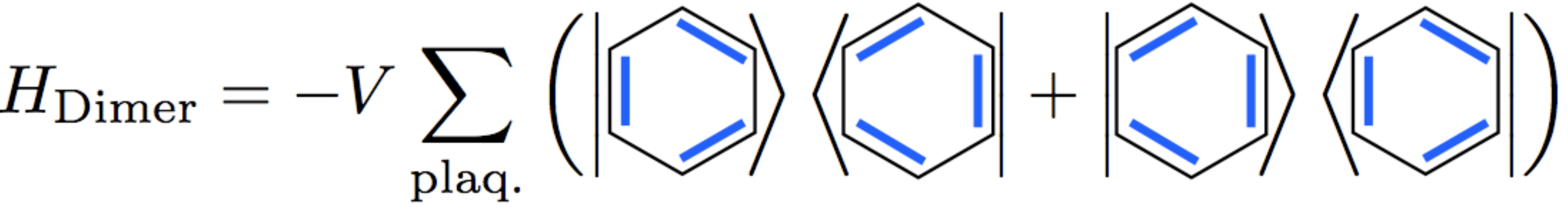}
\end{equation}
so the $V\sim O(J_{\perp}^{3}/J_{\parallel}^{2})$ effect of the hexagonal ring exchange is to cause dimers to resonate around hexagonal plaquettes on the diamond lattice.

\begin{figure}[h!]
\begin{center}
\includegraphics[width=9cm]{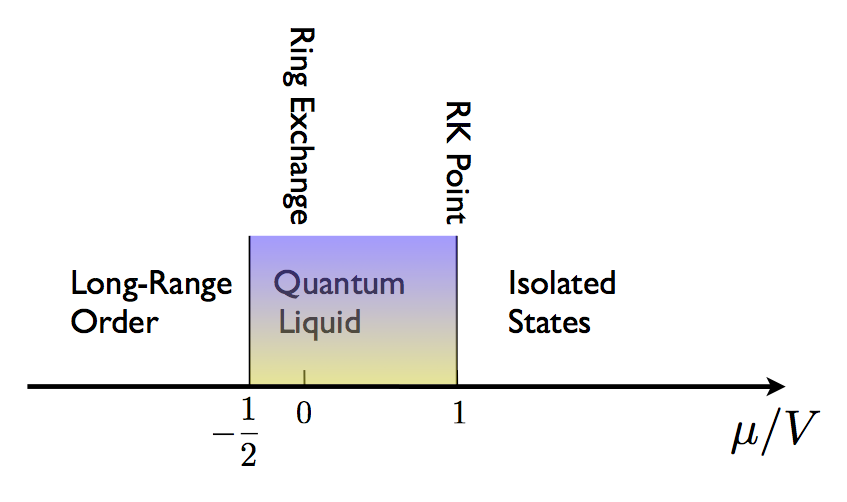}
\caption{Schematic phase diagram of the fully packed loop quantum dimer model described in the main text. \label{fig:PD}}
\end{center}
\end{figure}

It is possible to gain some insight into this model by adding the number operator for flippable plaquettes, $H_{\rm N}$, to $H_{\rm Dimer}$. The $H_{\rm N}$ term, which alone maximizes or minimizes the number of flippable hexagons depending on the sign of the coupling and allows one to tune the theory of quantum dimers to an exactly solvable Rokhsar-Kivelson (RK) point \cite{Hermele2004,Rokhsar}, is given by:
\begin{equation}
\label{eqn:dimer_model2}
\includegraphics*[width=0.6\linewidth]{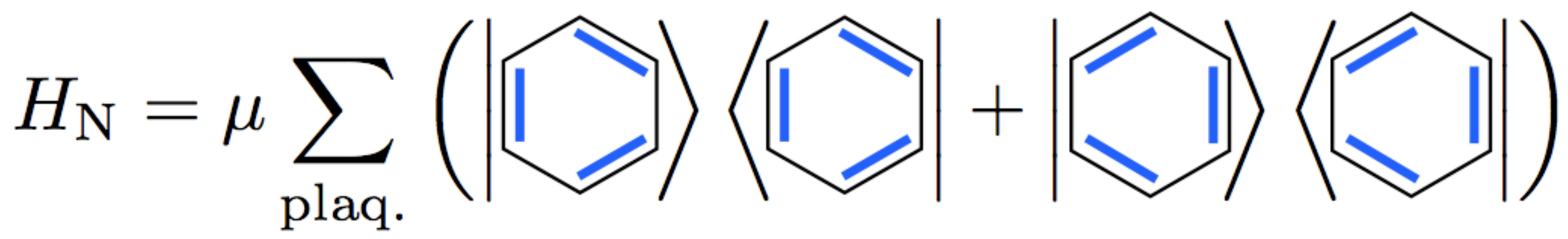}
\end{equation}
At the RK point $V=\mu$, one may write down the exact ground state wavefunction and obtain information about the excitations either through a single mode approximation \cite{Hermele2004,Moessner2003,Rokhsar} or by computing certain correlation functions in this model numerically exactly using classical Monte Carlo. The underlying insight in the latter procedure is that the ground state wavefunction of the RK model is an equal weight superposition of different ``loop coverings'' which can be sampled using Monte Carlo at infinite temperature \cite{Henley2004,LCA}. The result is that the low energy spectrum is gapless with a $k^2$ dispersion precisely at the RK point \cite{Hermele2004,Moessner2003}. This and other aspects of the RK point can be captured using an effective field theory which also allows one to infer the phase diagram in Fig.~\ref{fig:PD} of the model 
$H=H_{\rm Dimer} + H_{{\rm N }}$  \cite{Hermele2004,Moessner2003}. On one side of the RK point, $\mu>V$, perturbation theory tells us that the ground state immediately enters into a long-range ordered crystalline dimer phase. On the opposite side, we expect the number of flippable plaquettes to be maximized when $\mu$ is sufficiently large and negative $-$ producing
 another state with long-range order which has been named ``squiggle state" for the intertwined loops of dimers characterizing the phase
 \cite{Shannon2011}.
Directly away from the RK point, for $\mu \lesssim V$ where the resonating plaquette term $H_{\rm Dimer}$ is important, a liquid state should persist within some window of couplings $(\mu/V)_{\rm c}<(\mu/V)<1$. The presence of a linearly dispersing photon-like low-energy excitation mode within this window can be inferred from an effective field theory \cite{Hermele2004,Moessner2003}. Specifically, one finds that, while the dispersion is strictly quadratic at the RK gapless point,  the dispersion becomes linear for $\mu/V < 1$. This can be rationalized on the basis of an effective noncompact field theory:  a term $(\nabla\times\mathbf{A})^2$ which vanishes at the RK point is a relevant perturbation which immediately drives the system into the $U(1)$ phase away from the RK point  \cite{Hermele2004,Moessner2003}. 

More recently, concrete evidence for the presence of a $U(1)$ phase has come from numerics. Fortunately, the dimer Hamiltonian,
$H=H_{\rm Dimer}+H_{\rm N}$ of Eqns.~(\ref{eqn:dimer_model}) and (\ref{eqn:dimer_model2}) has no sign problem and can be studied using quantum Monte Carlo (QMC). In the guise of a hard core boson model on a diamond lattice with large nearest-neighbour repulsion, the dimer model also arises at low energies. Such a model has been simulated using QMC implemented  using the stochastic series expansion (SSE) method  \cite{Banerjee2008}. These simulations find a state with no superfluid order and no Bragg peaks in the structure factor, and there thus appears to be a liquid phase down to the lowest temperatures accessed by the simulations. In a recent zero temperature Monte Carlo study, Shannon and co-workers  mapped out the whole phase diagram of the dimer model \cite{Shannon2011}.
In particular, the authors of Ref.~[\onlinecite{Shannon2011}] found further evidence for a gapless mode with linear dispersion persisting across a finite $\mu/V$ window between the exactly solvable RK point $\mu/V=1$ and $\mu/V \approx -1/2$. This includes the crucial ``physical'' $\mu=0$ point 
corresponding to the effective ring exchange model Eqn.~(\ref{eqn:dimer_model}) derived via perturbation theory starting from the original spin Hamiltonian. The conclusion is that there is good evidence that the spin model with Hamiltonian Eqn.~(\ref{eq:ringexchange}) acting within the ice manifold has a quantum spin liquid ground state.

\subsection{From loops to a gauge theory}
 \label{sec:loops2gauge}
 
One can gain a great deal of physical insight into the quantum spin ice model by making a set of transformations from the dimer Hamiltonian Eqns.~(\ref{eqn:dimer_model}) and (\ref{eqn:dimer_model2}) to obtain a lattice gauge theory \cite{Hermele2004,Fradkin,CastroNeto2006}. These transformations require two main steps: the first is to enlarge the Hilbert space of the model by allowing the link dimer occupation numbers to take any integer value on each link ${\rm L}$, not only zero and one. The associated dimer occupation operator is $n_{\rm L} $. The original physical subspace of $S^{z}$ eigenvalues, $S_{\rm L}^{z}=\pm 1/2$, is a subset of  $S_{\rm L}^{z}= n_{\rm L}- 1/2$ with eigenvalues of $n_{\rm L}$  running over all integers. \footnote{This is not to say that we have increased the effective spin. Instead, the Hilbert space has been enlarged to that of a set of U(1) quantum rotors.} To recover the original physical subspace ($S^{z}=\pm 1/2$), which corresponds to the aforementioned hardcore dimer constraint, one introduces a ``soft'' constraint with tunable coupling through an extra term in the Hamiltonian of the form 
\begin{equation}   H_{\rm Constraint } =  U \sum_{\rm L} \left( n_{\rm L} - \frac{1}{2}  \right)^2.  \label{eqn:HConstraint}  \end{equation}
There is also a requirement for a second constraint which, taken together with the constraint on the occupation numbers on a link, imposes the ice-rule condition. This second constraint takes the form
\begin{equation}  Q_{\rm I} \equiv \sum_{\rm L \in Diamond sites} n_{\rm L} = 2 \label{eqn:constraint}  \end{equation}
where the sum runs over the four links connected to a given diamond site ${\rm I}$. The dimer constraint introduced in the previous section is that exactly two dimers connect to each diamond site. Now, the ``soft" constraint  of Eq.~(\ref{eqn:HConstraint}) together with $Q_{\rm I}=2$ enforce the ice-rule pattern of dimers. The constraint $Q_{\rm I}=2$ can be thought of as an analogue of Gauss' law. To see this, we assign an orientation to the diamond lattice links using the bipartiteness of that lattice as follows. The $A$ sites are those defined as ``UP" tetrahedra in Fig.~\ref{fig:pyrochlore} and the $B$ sites corresponding to the ``DOWN" tetrahedra. We let the links have orientation towards the $A$ sites and away from the $B$ sites. Having done this, we introduce so-called oriented link variables defined through
\begin{equation}
B_{{\rm L}_{A\rightarrow B} } =  +\left( n_{\rm L}- \frac{1}{2}\right) 
\end{equation}
which we call magnetic fields (since they are related to the orientation of the physical microscopic magnetic moments) and with the sign reversed when the orientation is reversed ($B_{{\rm L}_{A\rightarrow B} }=-B_{{\rm L}_{B\rightarrow A} }$). Now, the constraint on $Q_{\rm I}$ in Eq.~(\ref{eqn:constraint}) is 
\begin{equation} 
 \sum_{\rm L} B_{\rm L} = 0 ,
 \label{eqn:gauss}
 \end{equation}
which is recognizably Gauss' law discretized on a lattice. 

The ring exchange term of Eqn.~(\ref{eq:ringexchange}) gets modified when we enlarge the Hilbert space from the dimer model by moving over to the occupation number operators $n_{\rm L}$. The conjugate variables to $n_{\rm L}$ are phases $\phi_{\rm L}$ and the operators $\exp(\pm i \phi_{\rm L})$ raise ($+$) and lower ($-$) the occupation numbers on links. In the next step, to make the correspondence with a $U(1)$ gauge theory, we give these phases an orientation on the links (as described above) and rename them $\phi_{\rm L}\rightarrow A_{\rm L}$. With these steps, we obtain the Hamiltonian \cite{Hermele2004,Fradkin,CastroNeto2006}
\begin{equation}  H_{\rm Gauge} = U \sum_{\rm L \in Links} B_{\rm L}^{2} - K \sum_{\rm P} \cos \left( \osum_{L\in \left\{\hexagon\right\}} A_{\rm L}   \right)
\label{eqn:gaugetheory}    
\end{equation}
where the sum over $A_{\rm L}$ is an oriented sum around a hexagon and we call the flux variable through a given hexagonal plaquette ${\rm P}$ the electric flux $E_{\rm P}$. This electric flux is not to be confused with the physical (or fundamental) electric field which would enter the theory when considering, for example, the dielectric response of the system \cite{Saito2005}.

\begin{figure}[h!]
\begin{center}
\includegraphics[width=9cm]{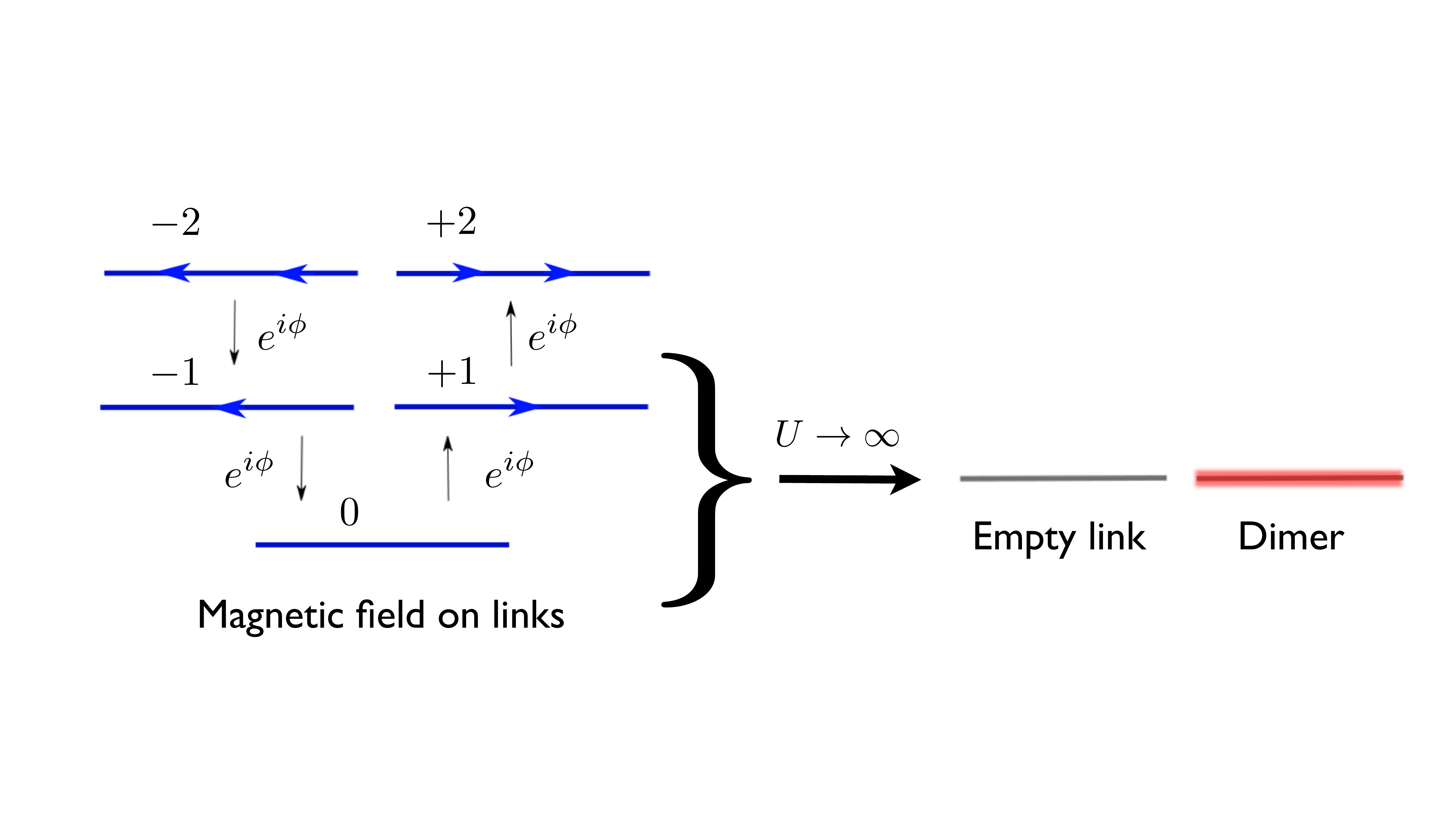}
\caption{Figure illustrating the connection between the gauge theory degrees of freedom and the quantum dimer states. Each link of the diamond lattice in the gauge theory has a tower of states, labelled by integer $n_{\rm L}$ connected by raising and lowering operators $e^{\pm i\phi_{\rm L}}$ where $\phi_{\rm L}$ is an operator with continuous spectrum.The $n_{\rm L}$ and $\phi_{\rm L}$ on each link are unoriented. Then they are assigned an orientation they become, respectively, magnetic fields $B_{\rm L}$ and vector potentials $A_{\rm L}$. When $U$ in Eq.~(\label{eq:HConstraint}) is taken to be large, all magnetic field states except for $B_{\rm L}=0,\pm1$ are gapped out. The low-lying magnetic field states coincide with the allowed dimer states with occupation number $n_{\rm L}=0$ and $n_{\rm L}=1$ on each link.  \label{fig:DimersFields}}
\end{center}
\end{figure}

In summary, the first term in the right hand side of this Hamiltonian is there to impose the constraint on the (diamond lattice) link occupation number $n_{\rm L}$ so that it assumes only values $0$ and $1$. The second term in Eqn.~(\ref{eqn:gaugetheory}) is the ring exchange of Eqn.~(\ref{eqn:dimer_model}), but now written explicitly in terms of a vector potential $A_{\rm L}$. The model is consistent when Gauss' law, Eqn.~(\ref{eqn:gauss}), is satisfied.
 This is a quantum theory because the field components satisfy canonical commutation relations $\left[ B_{\rm L} , A_{\rm L'}  \right] = -i\delta_{\rm LL'}$. Taken together, these ingredients constitute a version of quantum electromagnetism on the lattice with $B_{\rm L}$ as the magnetic field components and the cosine in the right hand term of the Hamiltonian (Eq.~\ref{eqn:gaugetheory}) as the lattice analogue of electric flux. Schematically, this route from the dimer model to a gauge theory is outlined in Fig.~\ref{fig:DimersFields}.
 
 We note that the vector potential $A_{\rm L}$ is defined modulo $2\pi$ in contrast to ordinary electromagnetism where it is defined over all real numbers. The effective low energy gauge theory of quantum spin ice is therefore a {\it compact} gauge theory with gauge group $U(1)$, the group of phase rotations. The origin of the compactness is the discrete spectrum of magnetic field states on each link which itself comes from the discrete dimer constraint. The compactness of the $U(1)$ gauge theory has one crucial consequence which sets it apart from ordinary electromagnetism, namely, that there is a novel charge-like excitation in the theory $-$ the vison $-$ with no correspondence to the magnetic monopoles of classical spin ice. This lowest energy gapped excitation plays (see Figure~\ref{fig:Excitations}) a crucial role in determining the phases of the lattice gauge theory.
 
Having reviewed the essential steps leading to the construction of a gauge theory for quantum spin ice systems, we discuss the phases of this theory. We will end up with the conclusion that the quantum liquid phase of the dimer model of the previous section can be thought of as the {\it deconfined phase} of the compact $U(1)$ gauge theory. This means that the compactness of the gauge theory is unimportant in the low energy limit or, equivalently, the fluctuations of the electric flux are small. In the deconfined phase, we may then expand the cosine of the Hamiltonian 
Eqn.~(\ref{eqn:gaugetheory}), omitting all but the lowest order nontrivial contribution. The Hamiltonian is then
\begin{equation}  H_{\rm Gauge, Deconfined} = U \sum_{\rm L \in Links} B_{\rm L}^{2} + K \sum_{\rm Plaq.} E^{2}_{\rm P}  
\label{eqn:deconfined}    
\end{equation}
where $E_{\rm P}$ is, slightly unconventionally (see Section~\ref{sec:conventions}),  the circulation of $A_{\rm L}$ around plaquette ${\rm P}$
\[
E_{\rm P} = \osum_{L\in \left\{\hexagon\right\}} A_{\rm L}.
\]
This Hamiltonian is recognizably a discretized form of the Hamiltonian describing electromagnetism in the absence of charges. This means that the excitations at low energy are photons $-$ with linear dispersion and two transverse polarizations.

Before we discuss in more detail the phases of Eqn.~(\ref{eqn:gaugetheory}), we take the opportunity to make the following observation. The discretized compact electrodynamics that we have discussed above differs from the 
standard Abelian lattice gauge theory that has been studied traditionally by lattice gauge theorists in one crucially important respect. Namely, at strong coupling $U/K\rightarrow \infty$, the Hamiltonian of Eqn.~(\ref{eqn:gaugetheory}), has 
 $n_{\rm L}=0,1$ within its vacuum (constrained by Gauss' law) corresponding to the spin ice configurations or superpositions of these states. Typically, pure Abelian gauge theories at strong coupling have a trivial vacuum with $n_{\rm L}=0$. In the literature, gauge theories with a trivial vacuum have come to be called ``even", or unfrustrated, gauge theories to contrast them with gauge theories such as Eqn.~(\ref{eqn:gaugetheory}) which are called  ``odd" or frustrated gauge theories \cite{Fradkin}. The nature of the vacuum states at strong coupling makes a dramatic difference to the nature of the phase diagram 
as a function of the coupling $K/U$. 

As discussed above, the unfrustrated gauge theory (with a trivial vacuum at large $U$) is known to have two phases in the $3+1$ dimensional case of interest here due to the well known work of Refs.~[\onlinecite{Polyakov,Polyakov1975,Polyakov1977,BMK,Guth}]. When $U/K$ is small, there is a gapless photon excitation and a Coulomb law between test charges (both monopoles and visons) inserted into the system. 
This is the deconfined, Maxwell or Coulomb phase. In the opposite limit, with large $U/K$, the theory is confining, meaning that the photon is gapped out and particles with opposite electric charges are bound together by a potential which grows linearly in the separation of the charges. There is a critical $(U/K)$ value at which the theory undergoes a transition between a confined and deconfined phase \cite{Polyakov,Polyakov1975,Polyakov1977}. 

The frustrated theory, in contrast, maps to the $\mu/V=0$ dimer model for large $U/K$. The evidence from numerical simulations is that the dimer model is in a deconfined phase when the ring exchange term is the only term present in the Hamiltonian (i.e. at the $\mu/V=0$ of the phase diagram shown in Fig.~\ref{fig:PD}) \cite{Shannon2011,Banerjee2008}. When $U/K$ is small in the frustrated gauge theory, the frustration should not be important and, once again, we expect the theory to be deconfined. In summary, it appears that confinement may be completely absent in the frustrated gauge theory, though, to our knowledge, this has not been established beyond the heuristic arguments given here. Since the deconfined phase of a $U(1)$ gauge theory exhibits universal features, it should not matter whether the dimer model (in its deconfined phase) maps to a frustrated or unfrustrated gauge theory $-$ the excitations below the energy scale of the hard core dimer violating fields (set by $U$) in the frustrated gauge theory can be inferred from the unfrustrated gauge theory about which much is known \cite{Polyakov,Polyakov1975,Polyakov1975,BMK,Guth}. In the next section, we borrow insights from the deconfined phase of the unfrustrated gauge theory to say something about the phenomenology of the deconfined phase of the quantum dimer model or, by extension via the series of mappings reviewed above, quantum spin ice.

\subsection{Quantum liquid: excitations and phenomenology}
\label{sec:phenom}

The perspective on the quantum spin ice problem gained by thinking about the deconfined phase of $U(1)$ lattice gauge theory allows us to infer the nature of the excitations in the dimer model. At energies below the scale of the ring exchange $\mathcal{J}_{\rm ring}$ in the deconfined phase, where fluctuations of the gauge field $A_{\rm L}$ are small, the compactness of the theory should not matter, and we recover familiar noncompact electromagnetism with gapless photon excitations with a pair of transverse components. 

At much higher energies, on the scale of the Ising exchange term $J_{\parallel}$ in the original spin model, there are gapped magnetic charge excitations 
(the ``monopoles" of Section~\ref{sec:spin_ice})
corresponding to spin flips out of the manifold of spin ice states. Both in classical and in quantum spin ice, single spin flips correspond to a breaking of the divergence-free condition on the magnetic field, $B_{\rm L}$, and hence the creation of magnetic field sources or magnetic charges. Via successive spin flips, it is possible to separate these charges. 
Whereas in classical spin ice an effective magnetostatic interaction arises between these charges owing to thermal averaging solely over circulations of the field $B_{\rm L}$, in quantum spin ice these magnetic charges interact both with the emergent magnetic {\it and} electric fields. \footnote{In this paragraph we refer to nearest neighbour classical spin ice in which there is an emergent Coulomb potential between monopoles which is of entropic origin. In dipolar spin ice, there is an effective Coulomb potential arising directly from the dipolar interaction.}

At intermediate energies, a third type of excitation is present which arises from the compactness of the gauge field. 
These are the aforementioned visons. For readers familiar with $Z_2$ spin liquids, it might be helpful to point out that the magnetic charges in $U(1)$ spin liquids are the analogues of so-called gapped $Z_2$ flux excitations also called visons in these systems \cite{Senthil}. One way of seeing that these should be present is as follows. The loops of dimers in the quantum dimer model introduced above may be interpreted as magnetic field strings within the language of the gauge theory. 
 While the dimer model is defined to be a theory of closed strings, excitations out of the spin ice manifold may occur in the original spin model breaking these strings and lead to the magnetic charges (monopoles). Importantly, there are now also electric loops in the theory. These,  unlike the magnetic strings, are not imposed by a kinematic constraint of Eqn.~(\ref{eqn:constraint}). Instead, they arise from the dynamics of the dimer model -- the resonating hexagonal plaquettes form loops of electric fluxes. When these flux strings break, the string endpoints form new sources: electric charges which are gapped in the liquid phase with a gap of the order of 
${J}_{\rm ring}$. This picture of visons appearing at ends of broken (electric) strings is true also for the vison excitations in $Z_2$ spin liquids. The visons are massive particle-like excitations with a net charge. Given the pyrochlore lattice structure, they can be thought of as hopping on a second diamond lattice displaced from the original diamond lattice on which the magnetic charges hop by half of one elementary cubic cell in each coordinate direction. ~\footnote{It is worth pointing out that magnetic monopoles in high energy physics have a similar origin to the vison excitation discussed here. In brief, gauge theories where electromagnetism arises from a compact $U(1)$ subgroup of a larger compact gauge group have gapped magnetic charge excitations which are generalisations of the visons obtained here. In the Standard Model they do not appear because the electromagnetic gauge group is non compact. However, magnetic monopoles appear when the Standard Model is embedded in a theory with a larger symmetry including the famous $SO(10)$ Grand Unified Theory of the strong, weak and electromagnetic interactions \cite{GUT}. One of the features of compact $U(1)$ gauge theory is that charges (the visons and monopoles) are naturally quantised. This feature assumes some importance in fundamental physics where the quantisation of charges is something that one would like to explain. Within the aforementioned GUT, the appearance of monopoles goes hand-in-hand with the quantisation of the fundamental charges (among other important features). Such GUTs have inspired experimental searches for magnetic charges \cite{MMs}.}

Having noted the types of excitations and the hierarchy of energy scales in quantum spin ice, we turn to plausible
experimental signatures of a magnet with a low energy $U(1)$ phase. At the highest temperatures, the magnet is a featureless paramagnet. Upon cooling, the material enters a classical spin ice regime with a temperature dependent density of monopoles and a Pauling residual entropy. This state of matter is well-known to exhibit distinctive dipolar spin-spin correlations which show up as pinch points in the neutron scattering cross section  \cite{Henley2010}. The quantum dimer model at infinite temperature is nothing other than classical spin ice and, therefore, this model does not capture the crossover into the true high temperature paramagnetic regime of the underlying spin model 
 \cite{Benton2012}. 

\begin{figure}[h!]
\begin{center}
\subfigure[Inelastic Scattering]{
\includegraphics[width=12cm]{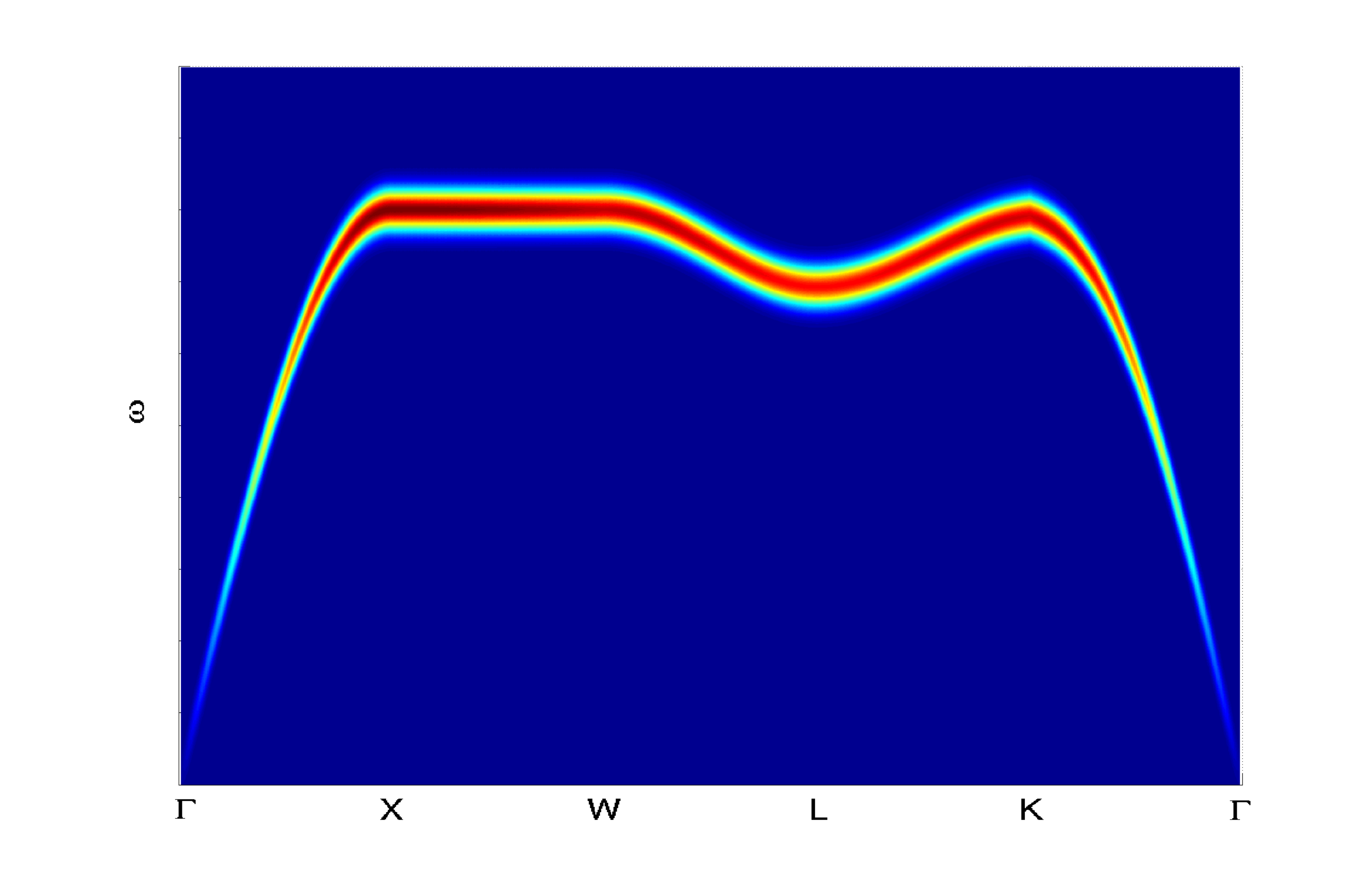}
\label{INS}
}
\subfigure[Energy Integrated. Zero T]{
\includegraphics[width=7cm]{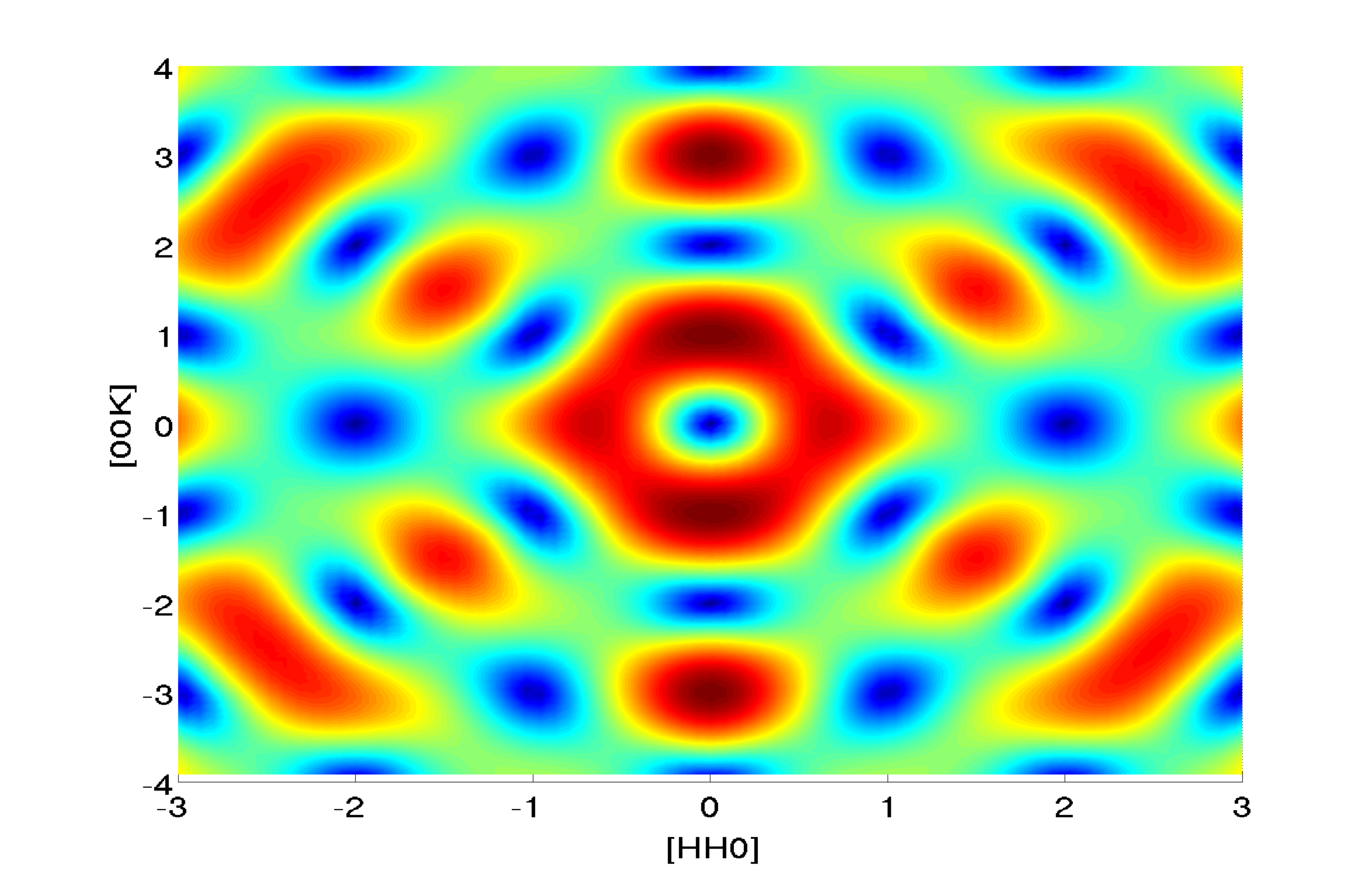}
\label{ET0}
}
\subfigure[Energy Integrated. $T=10ca_{0}^{-1}$]{
\includegraphics[width=7cm]{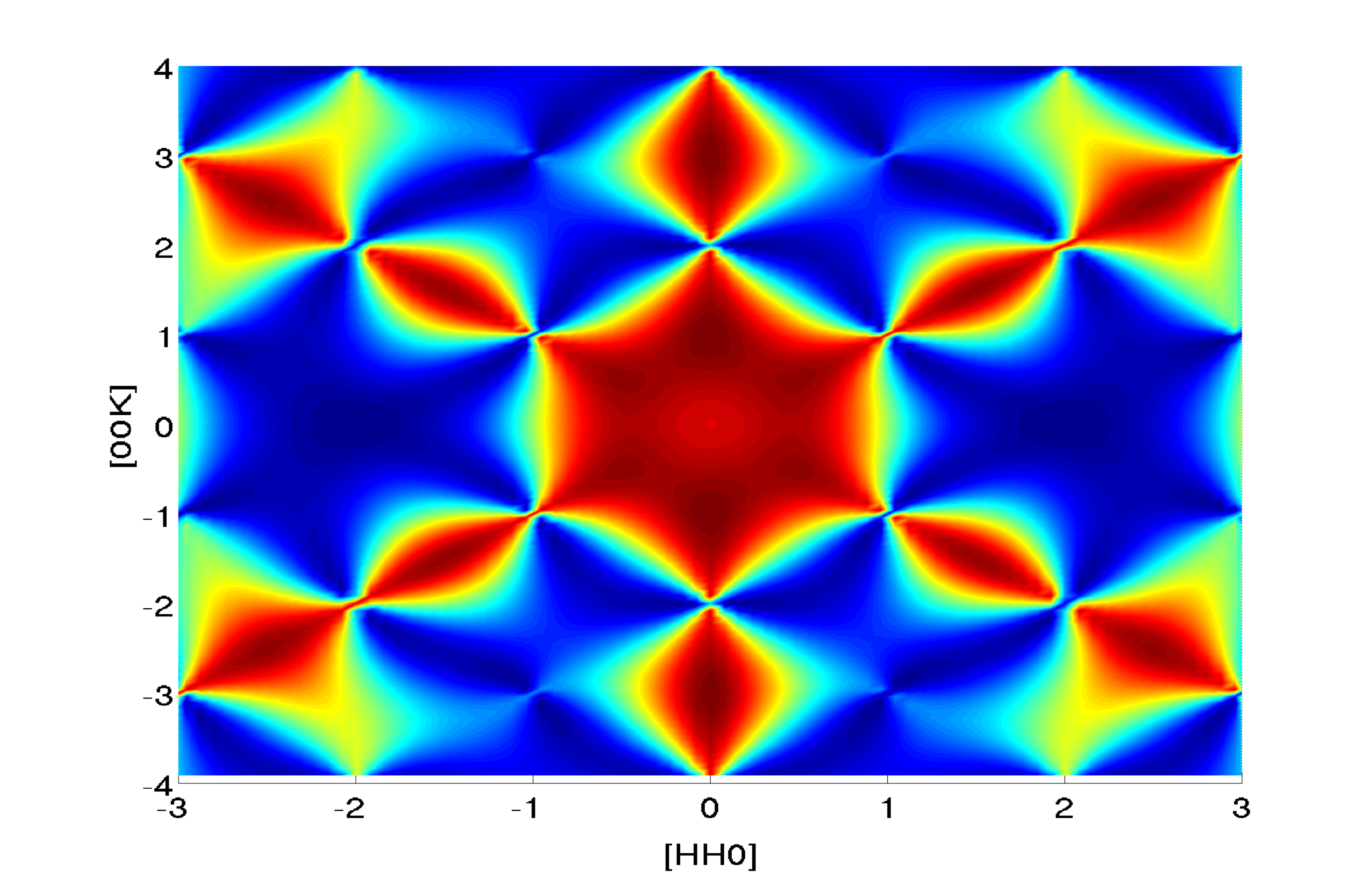}
\label{ET10}
}
\end{center}
\caption{Characteristic neutron scattering patterns observed for quantum spin ice. Panel (a) shows the characteristic inelastic scattering from photon excitations. The reciprocal space path is taken along straight lines between high symmetry points in the Brillouin zone. Panels (b) and (c) show the 
unpolarized energy integrated scattering for photon scattering for zero temperature and high temperature (compared to the ring exchange) respectively. The high temperature plot is similar to the scattering expected for classical spin ice. The temperature is measured in units of $c/a_{0}$ where $c=\sqrt{UK}a_{0}\hbar^{-1}$ is the velocity of the linearly dispersing excitations and $a_{0}$ is the cubic unit cell edge length for the pyrochlore lattice. 
The dynamical structure factor and unpolarized energy-integrated scattering were computed from formulae in Ref.~[\onlinecite{Benton2012}]. 
\label{fig:neutron}}
\end{figure}

At the lowest temperatures, within the $U(1)$ phase, there should be quite distinctive experimental signatures. While there should be no magnetic Bragg peaks, inelastic neutron scattering can in principle probe the linearly dispersing photon modes as recently worked out in Ref.~[\onlinecite{Benton2012}] since, like magnons, these excitations carry spin one. Our calculation of the expected inelastic 
scattering pattern (based on that in Ref.~[\onlinecite{Benton2012})] between high symmetry points in the Brillouin zone is shown in Fig.~\ref{INS} where the most energetic modes appear on the scale of the ring exchange coupling. The energy integrated scattering at zero temperature is presented in 
Fig.~\ref{ET0}. Thermal fluctuations cause the photons to decohere and the neutron cross section crosses over smoothly into the pinch point scattering of classical spin ice \cite{Benton2012}.

At intermediate temperatures, the phenomenology is, to date, not completely clear. The $T^3$ (thermal radiation) law in the heat capacity should break down as visons and magnetic  charges become thermally nucleated and the residual entropy recovers from zero, at zero temperature, to the Pauling entropy, in the classical ice regime. These effective charges will also lead to diffuse neutron scattering as is well-known in classical spin ice \cite{Henley2005,Sen}. 

 Also, the analogy with the compact $U(1)$ gauge theory suggests that thermal fluctuations  should give the photon a small mass and the otherwise Coulomb-like electric and magnetic charges should become screened. \footnote{At finite temperature, the temporal direction of the $3+1$ dimensional gauge theory is wrapped onto a circle of radius $2\pi\beta$ where $\beta$ is the inverse temperature. At high temperatures the radius of the circle is small so that the $3+1$ dimensional gauge theory may be viewed as a $2+1$ dimensional gauge theory at zero temperature. In three dimensions, the Wilson loop expectation value exhibits an area law \cite{Polyakov}. Then, the same argument that implies that the $2+1$ dimensional gauge theory has a gapped photon can be used to see that the four dimensional gauge theory does as well at high temperatures  \cite{PolyakovFT,SusskindFT}}. Whereas the thermal occupation of photon modes is sufficient to observe the crossover between quantum and classical ice, and although the quantum liquid at zero temperature is adiabatically connected to the trivial paramagnet,  a more featured scenario is possible. In particular, a form of gauge mean field theory for quantum spin ice produces, at finite temperature, a truly novel behavior: a first order transition between the high temperature paramagnet and the quantum spin ice state \cite{SavaryFT}.

\subsection{Stability of the $U(1)$ liquid}
\label{sec:stability}

One of the remarkable features of the $U(1)$ spin liquid is that it is stable to all local perturbations \cite{Hermele2004,Foerster1980,Wen}. This is surprising for at least two reasons. One is simply that the phase is gapless and there is therefore an {\it a priori} danger that some perturbations may open a gap. A second reason is that the gauge invariance of the lattice model is not exact but is an emergent property at low energies. One can see by power-counting that all gauge noninvariant perturbations to the Maxwell action are relevant in the renormalization group (RG) sense in (3+1) dimensions \cite{Weinberg_I}  so one might expect that the $U(1)$ liquid would not survive such perturbations. \footnote{The deconfined phase comes with a natural cutoff -- the magnitude of the transverse terms in the spin Hamiltonian -- below which we expect that we do not have to worry about gauge invariant interactions beyond the Maxwell action. One can see that this is the case by studying the kinds of gauge invariant terms that one can add to the low energy effective theory. The aptly named irrelevant operators in the theory are suppressed by powers of the (large) cutoff at long wavelengths so that their effect at sufficiently low energies is negligible. The relevant operators are not suppressed in this manner. Power counting is a simple criterion to assess how couplings flow under a change of scale. Specifically, the irrelevant operators have couplings with dimensions of some negative power of the energy scale where the dimension is determined directly from the  action.} Crucially, for the search of this exotic state of matter among real materials, it turns out that this fear is likely not justified as we now discuss.

Because the $U(1)$ gauge theory comes from a microscopic spin model on a lattice, the gauge group is compact so gauge non-invariant terms such as $M^{2}\mathbf{A}^{2}$ are not allowed in the theory. Instead, the gauge non-invariant perturbations can only appear through terms like $\exp(iA)$ which are magnetic monopole hopping operators, or from terms that hop visons. But then, since both 
{\it matter field} (vison and magnetic) excitations are gapped, these gauge non-invariant operators cannot affect the low energy physics $-$ the $U(1)$ liquid is stable to all gauge non-invariant perturbations $-$ while all gauge invariant perturbations are irrelevant in the RG sense. There is some work putting these arguments on a rigorous footing using the idea of quasi-adiabatic continuity. The idea, roughly stated, is to switch on a gauge non-invariant perturbation to the $U(1)$ liquid with small coupling $s$ which has the effect of taking the ground state $\vert \Psi (s=0)\rangle$ into $\vert\Psi(s)\rangle$.  Then one transforms all the operators $O$ in the theory in tandem with the switching on of the perturbation in such a way as to  (i) preserve the locality of the operators and (ii) so that the expectation values of operators $O(s)$ computed using the $\vert\Psi(s)\rangle$ and the dressed operators are the same as those computed using $\vert \Psi (s=0)\rangle$ and the undressed operators $O$, up to corrections that vanish in the thermodynamic limit. The authors of  Ref. [\onlinecite{Hastings2005}] were able to show that, although the bare Hamiltonian has small gauge non-invariant terms, the generators of gauge transformations are dressed after the addition of these perturbations to the bare model in such a way that a local gauge invariance survives in the dressed model. Although unproven, this is thought to preserve also the gaplessness of the theory.

\subsection{Naturalness of the $U(1)$ liquid}
\label{sec:natural}

In the previous section, we explained that the $U(1)$ liquid has the remarkable property of being stable to all local perturbations. This means that if the $U(1)$ liquid is known to occur at some point in the space of all possible couplings, then one can vary the couplings in any direction in the space of couplings by a finite amount with the ground state remaining in this phase (unless our original point is at a phase boundary). 

While it might be possible to engineer a quantum system, perhaps a trapped cold atom system, that enters the $U(1)$ phase, in the immediate future the most likely candidate systems that might host such a phase are certain pyrochlore magnets. Assuming this to be the case, the stability of the $U(1)$ liquid is of marginal relevance compared to the broader issue of whether the space of couplings explored by real materials accommodates the $U(1)$ phase over a significant region in this space. In short, it is useful to know whether the $U(1)$ liquid is too finely tuned to be observed in at least one of the set of available materials. If a $U(1)$ liquid were discovered we could, with reasonable confidence, pronounce that such phases are natural in parameter space. Since a deconfined $U(1)$ liquid phase has not been found in any of the pyrochlore magnets (at the time of writing), the question is worth asking from the perspective of microscopic models because it has a bearing on the discoverability of such phases and may provide some guidance to the experimental search. 

\begin{figure}[htp]
\begin{center}
\subfigure[Kramers]{
\includegraphics[width=9cm]{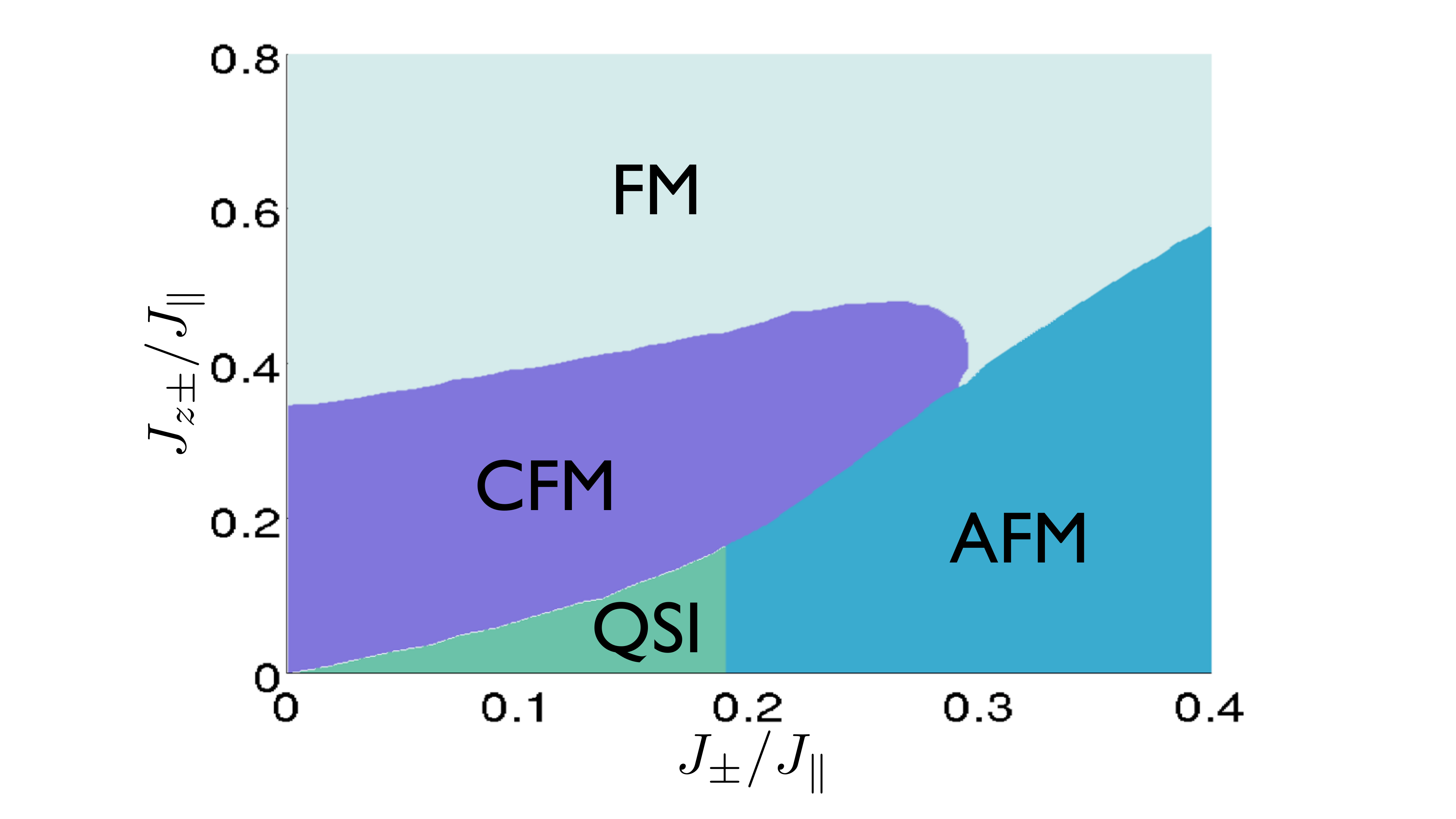}
\label{Kramers}
}
\subfigure[Kramers. Corrected.]{
\includegraphics[width=9cm]{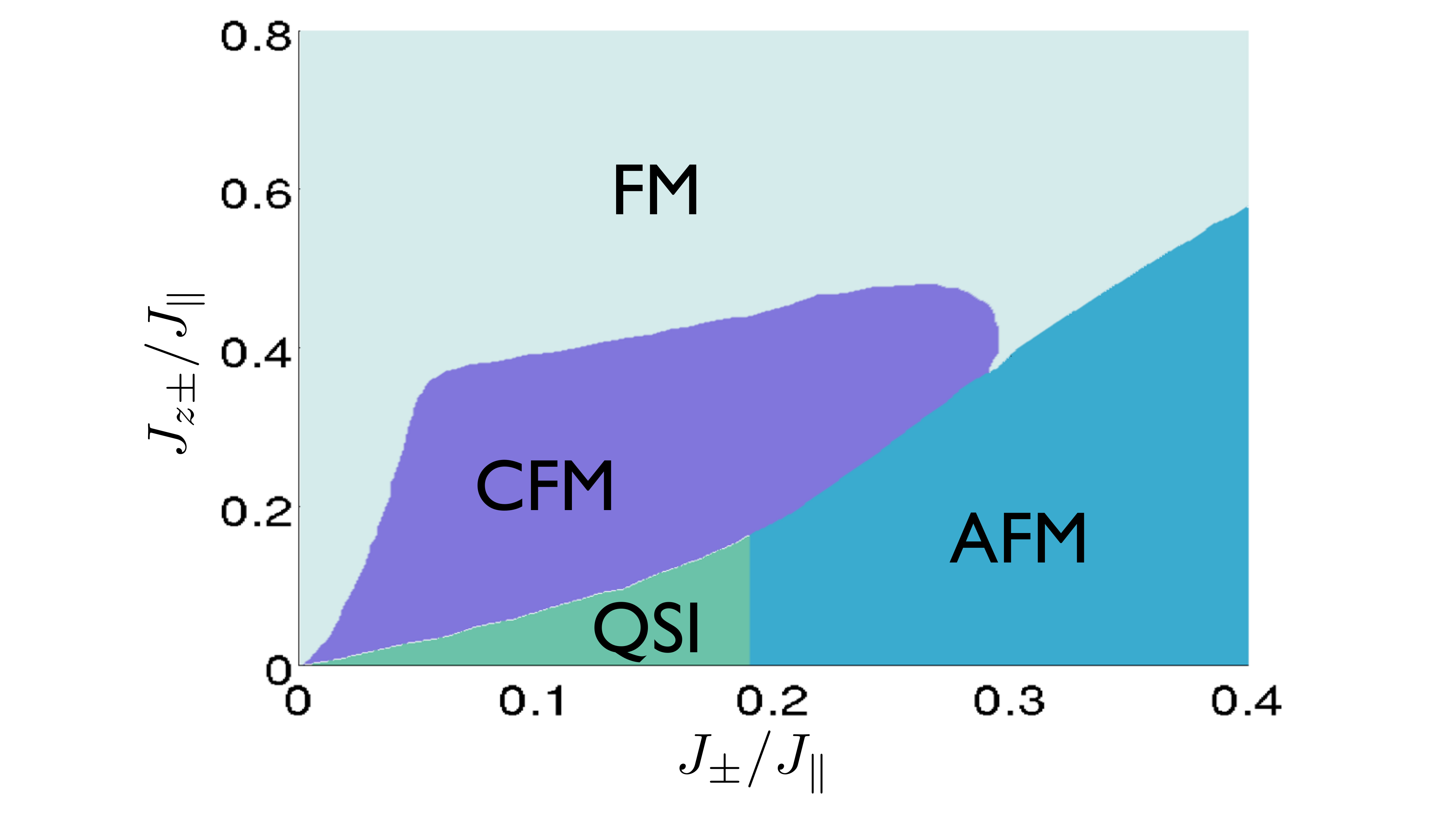}
\label{Kramers with perturbative correction}
}
\subfigure[Non-Kramers]{
\includegraphics[width=9cm]{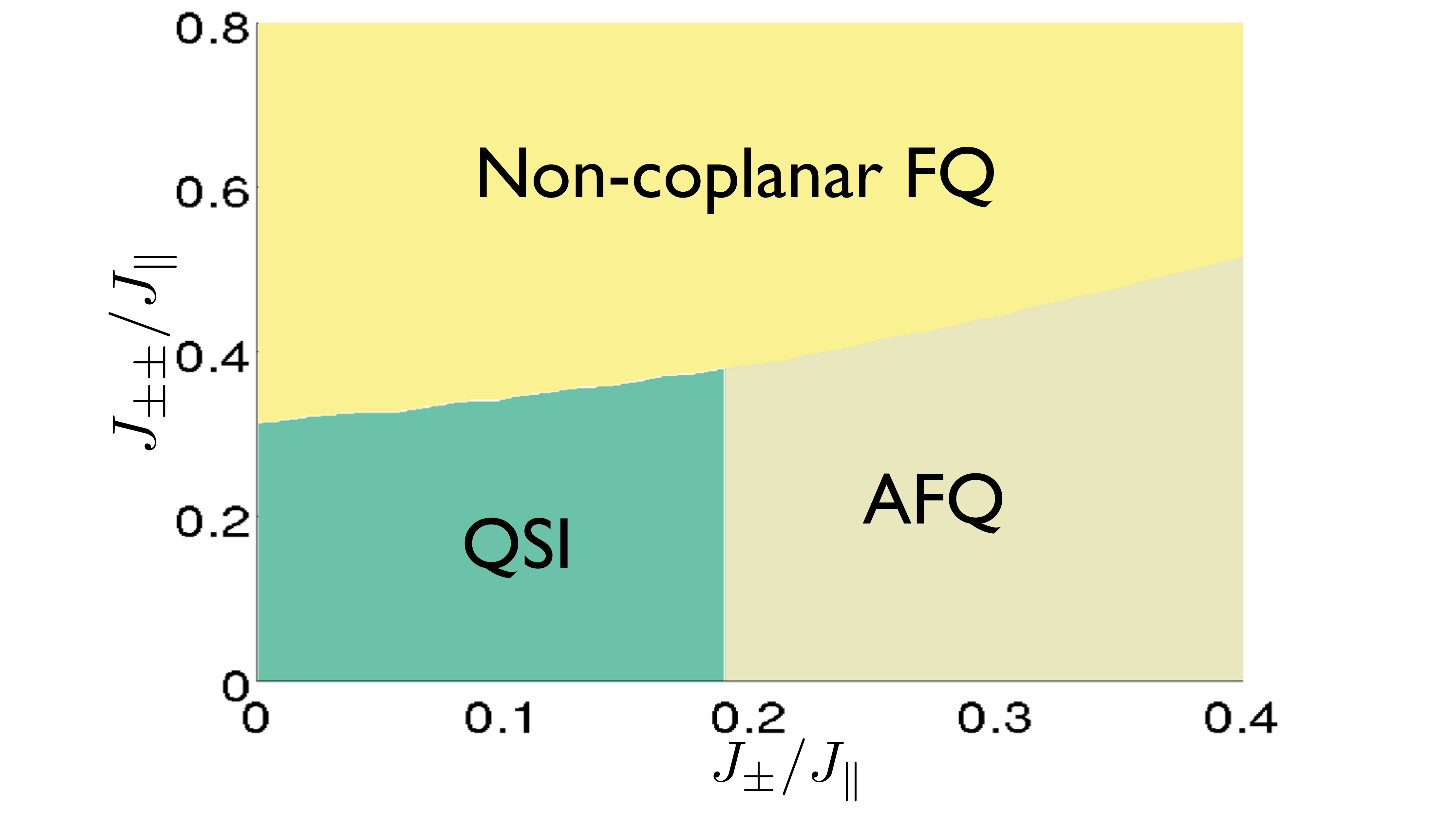}
\label{Non-Kramers}
}
\end{center}
\caption{Zero temperature mean field phase diagrams taken through sections in the space of symmetry-allowed nearest neighbour exchange couplings on the pyrochlore lattice. The mean field theory (gMFT), described in the main text, can capture both conventionally ordered and quantum spin liquid phases. The top panel (for odd electron magnets) shows a pair of exotic phases emerging from the classical spin ice point $J_{\pm}=0$ and $J_{{\rm z}\pm}=0$ - the quantum spin ice (QSI) and Coulombic ferromagnet (CFM).  The gMFT does not capture the perturbatively exact phase (in small parameter $J_{{\rm z}\pm}/J_{\rm zz}$) \cite{Savary2011,SavaryFT}. Panel (b) is a schematic plot to show the likely effect of correcting the gMFT to account for the perturbative result. Panel (c) is a phase diagram for even electron systems showing the quantum spin ice phase and two quadrupolar phases \cite{Lee2012}.
\label{fig:gMFT}}
\end{figure}

To date, the naturalness of quantum spin liquids in general is largely an open question \cite{Balents_Nature}. For the particular problem of the naturalness of the Coulomb phase in pyrochlore magnets, the location of the $U(1)$ liquid phase within the space of nearest-neighbour anistropic exchange couplings has been partially mapped out within a form of gauge mean-field theory for even electron (Kramers) magnetic ions \cite{Savary2011} and odd electron (non-Kramers) ions \cite{Lee2012}. The zero temperature gauge mean-field theory (gMFT) employed by the authors of these papers is a variant of slave particle mean field theory which, in this case, involves the formally exact step of splitting the anisotropic exchange Hamiltonian Eqn.~(\ref{eq:Heff}) 
in Section \ref{sec:materials_considerations} below
into gauge field degrees of freedom defined on links of the diamond lattice and new boson fields defined on the centres of diamond sites. The latter bosonic fields are referred to as 
{\it spinons} in that paper and as magnetic monopoles here (see Section \ref{sec:conventions} on naming conventions).
One decouples the resulting interacting theory and solves self-consistently for the expectation values of the gauge field and monopole fields. Whereas, the frustrated gauge theory discussed in Section~\ref{sec:loops2gauge} describes the physics of the quantum spin liquid within the spin ice manifold, the aim of the gauge mean-field theory is to capture further aspects of the physics of the full anisotropic spin model on the pyrochlore. This is the reason why, in addition to a compact gauge theory, one has couplings to electrically charged matter fields and one can expect Higgs phases, in addition to possible deconfined and confined phases. Higgs phases have the property that the photon is gapped out by the condensation of the bosonic matter fields; in some sense they are superconducting phases. The criteria for distinguishing different phases of the resulting gauge theory $-$ deconfined, confined and Higgs phases $-$ are reminiscent of those in works on the mean-field theory of lattice gauge theories (see for example, Refs.~[\onlinecite{BDI,Drouffe82}]). The principal difference between the types of theories considered in that early work compared to those arising from the pyrochlore spin models is that the former have explicit gauge kinetic terms in the action whereas, in the latter, there are only matter-gauge couplings $-$ the gauge
kinetic terms being solely generated by the dynamics of the monopole fields. Secondly, the phase diagram of the gauge theory can be interpreted in the language of the microscopic spin/magnetic degrees of freedom. Specifically, the confined phase corresponds to some ordered spin ice phase while Higgs phases are long-range
ordered magnetic phases in which 
the moments have nonzero expectation value perpendicular to the local Ising directions. Phase diagrams produced by solving the gMFT are shown in Fig.~\ref{fig:gMFT}.


The phase diagrams arising from the gauge mean-field theory show the $U(1)$ liquid phase surviving out to couplings $J_{\perp}/J_{\parallel}
\sim O(10^{-1})$ away from the classical spin ice point (where only $J_{\parallel}\neq 0$) \cite{Savary2011,Lee2012}. 
Supposing that the mean-field theory correctly captures the $U(1)$ phase boundary, we next turn to the naturalness of materials with exchange
couplings satisfying $J_{\perp}/J_{\parallel}\sim O(10^{-1})$. In other words, we consider the physics that leads to XXZ-like models in pyrochlore magnets 
conceptually akin to that of Eqn.~(\ref{eq:QSI}) with a dominant Ising term $H_{\rm CSI}$.

The crystal field, in tandem with spin-orbit coupling, which are responsible for the single ion magnetic anisotropy, become increasingly important for magnetic ions further down in the periodic table. This trend coincides with a reduction of the typical exchange couplings. Ising magnetism protected by the largest anisotropy gaps relative to the interactions is expected to occur among the rare earth or actinide magnets. Indeed, among the rare-earth pyrochlores, typical crystal field anisotropy gaps are of $O(10^2)$ K \cite{Rosenkranz,Bertin2012} while,
thanks to the trigonal symmetry, 
the single ion ground states are often doublets with effective exchange couplings of order $O(1)$ K between effective spin-1/2 degrees of freedom.

The crystal field ground state doublet of ``would be non-interacting'' non-Kramers ions is strictly Ising-like. In these systems,
effective transverse exchange couplings between the low energy effective spin one-half moments may arise in two ways: 
via multipolar couplings  \cite{Onoda2011,OnodaTanaka2011} and via mixing with excited crystal field levels  \cite{Molavian2007,Molavian2009}. 
The latter effect is suppressed by powers of $1/\Delta$, where $\Delta$ is the energy gap to the first excited crystal field states, 
while multipolar couplings 
of superexchange origin are suppressed by powers of the charge transfer gap. 
In such non-Kramers materials, Ising models with weak transverse couplings are therefore expected to be natural,
 as we discuss in Section \ref{sec:materials_considerations}. Pragmatically speaking, this is borne out 
by the existence of the spin ice materials which are very well described by a {\it classical} dipolar Ising spin ice model  \cite{denHertog,Yavorskii,Lin2013}. 
While the energy scale $J_{\perp}/J_{\parallel}$ is sufficiently 
small that the effective ring exchange term may be typically negligible, 
a number of rare-earth pyrochlores are coming to light that exhibit some quantum dynamics  \cite{Zhou2008,Wen_Pr2Zr2O7}.
We return to this topic in Section \ref{sec:candidate_materials}.
It is of considerable interest to establish the nature of the low temperature magnetism in these materials. Unfortunately, 
the degeneracy of non-Kramers crystal field doublets being accidental, it is sensitive to perturbations that need not 
be time-reversal symmetry invariant. It is then important to address the size of likely ring exchange terms compared to 
crystal field degeneracy-breaking perturbations.

The crystal field ground state doublets of Kramers ions are, by comparison, robust to time reversal invariant perturbations and there are no
a priori symmetry 
constraints on the strength of the relative strength of the effective anisotropic exchange couplings (such as $J_\parallel$ and $J_\perp$).
 Such materials are thus expected to afford an exploration of the full space of symmetry-allowed nearest-neighbour couplings. We would then expect
to commonly find $J_{\perp}/J_{\parallel}\sim O(1)$ among Kramers ions. 
The recently determined couplings in Yb$_{2}$Ti$_{2}$O$_{7}$ \cite{Ross2011,Applegate2012,Hayre2012} and Er$_{2}$Ti$_{2}$O$_{7}$ \cite{Savary_ETO,Oitmaa2013} 
bear out this expectation. 
The gMFT described above yields a region of parameter space in which the $U(1)$ liquid lives where the transverse terms 
are not necessarily much smaller than $J_{\parallel}$, indicating that the $U(1)$ liquid may not be unnatural in Kramers rare earth pyrochlores. 
Indeed, there is extensive on-going work exploring the nature of the low temperature phase of Yb$_{2}$Ti$_{2}$O$_{7}$ 
 \cite{Ross2011,Applegate2012,Hayre2012,Chang2012,DdR2012,DOrtenzio2013},
a matter we return to in the Section \ref{sec:Yb2Ti2O7}.

The perturbation theory deployed to find the ring exchange Hamiltonian earlier in Section~\ref{sec:spin2loops} does nothing to suppress 
Ising couplings beyond nearest neighbour  \cite{Ruff_PRL,Yavorskii} 
which may be of superexchange origin or from the long-range dipolar interaction  \cite{denHertog,Yavorskii}, 
the latter being typically large among rare-earth and actinide magnetic ions.  
Such Ising couplings tend to lift the degeneracy of the ice states  \cite{Melko2001,Melko2004} and the stability of the $U(1)$ liquid ultimately 
boils down, roughly, to a comparison of the energy scales of ring exchange and of the further neighbour Ising couplings
of the form $J_{ij}^{zz}(r_{ij}) S_i^{z_i} S_j^{z_j}$. Ref.~[\onlinecite{Benton2012}] includes a study of the effect of the third neighbour Ising coupling on quantum spin ice finding that it drives long range order above some threshold which is O(1) times the ring exchange coupling. It would be interesting to consider further the role of couplings beyond nearest neighbour on the $U(1)$ phase.

In summary, the lanthanide (4f) and actinide (5f) pyrochlore magnets offer a tantalising opportunity to discover quantum spin ice phases. For Kramers magnets among these materials, it is possible to satisfy all the criteria necessary to see quantum spin ice physics and the main difficulty to be overcome is the large space of parameters that such materials can, in principle, explore. In non-Kramers magnets, the pessimistic perspective that such materials should be of no interest for exotic quantum states of matter, because of their  large angular momentum $J$ and large single ion anisotropy is not correct. Thanks to deep low-energy Ising doublets, the underlying high-energy microscopic Hamiltonian can, once projected in the low-energy Hilbert space spanned by the doublets, be described by an effective spin-1/2 Hamiltonian allowing for significant quantum dynamics. Indeed, a microscopic calculation for a toy-model of  Tb$_2$Ti$_2$O$_7$ makes this point crisp \cite{Molavian2007,Molavian2009}. Unlike Kramers magnets, crystal field degeneracies in non-Kramers magnets is sensitive to being broken by disorder and sample quality will be a particularly important issue. Unfortunately, short of having  ab-initio calculations guidance, it is a matter of luck  finding the right material that falls in a regime of interactions where a $U(1)$ spin liquid is realized, or some other novel quantum states  \cite{Savary2011}. We discuss this topic in Section~\ref{sec:materials_considerations}.

\subsection{A broader perspective on quantum spin ice} 
\label{sec:context}

In this review, we have concentrated on the possibility that the deconfined phase of a $U(1)$ gauge 
theory can arise in certain pyrochlore magnets. This discussion would be incomplete without widening the 
scope a little by mentioning some other models with emergent low energy $U(1)$ liquid phases. 

A natural place to begin is the work of Baskaran and Anderson  \cite{Baskaran1988} and Affleck and Marston \cite{Affleck1988} who used slave bosons to study certain quantum spin models in $2+1$ dimensions. Within this approach, the spins are fractionalized into bosons with an accompanying $U(1)$ gauge redundancy. The pure $U(1)$ gauge theory is compact and is confining in $(1+1)$ and $(2+1)$ dimensions: the gauge boson is gapped out and the matter fields to which it is coupled are bound into states of zero gauge charge \cite{Polyakov,Polyakov1975,Polyakov1977}. This argument implies that the analogue of quantum spin ice in two dimensional magnets is  in a long-range ordered magnetic phase which has been explored in various works \cite{Chakravarty2002,Shannon2004,Syljuasen2006,Chern2013}. The implication of confinement for the slave boson procedure is that the fractionalization of spins is not a correct description of the physics in the cases studied in Refs.~[\onlinecite{Baskaran1988, Affleck1988}]. 

Fluctuations of the gauge field in slave particle descriptions of quantum spin models can be suppressed by taking the $SU(2)$ magnet to $SU(N)$ in the large $N$ limit leading to deconfinement even in two dimensional magnets. In this limit, slave particle mean field theory is a controlled approximation \cite{Wen,Wen2002a,Wen2002b}. A promising direction, motivated by slave fermion treatments of quantum magnets, comes from recent work showing that fractionalization is possible in $2+1$ dimensions either when $U(1)$ gauge fields are coupled to a number $n$ of different types of fermion with a Dirac dispersion \cite{Hermele2004b,Nogueira2005} (where $n$ need not be very large) or 
in the presence of a Fermi surface \cite{SSLee2008}. The physical significance of these results, and perhaps a broader lesson, is that a deconfined $U(1)$ liquid phase may arise in condensed matter models of {\it real} materials with two dimensional magnetism despite the fact that the minimal $2+1$ gauge-only $U(1)$ theory is confining \cite{Polyakov,Polyakov1975,Polyakov1977}. 

In $3+1$ dimensions, as we have described in preceding sections, the $U(1)$ gauge theory can have a deconfined phase. One of the earliest examples of a bosonic model with an emergent low energy electromagnetism is described in Ref.~[\onlinecite{Foerster1980}] which argues that certain compact but non gauge invariant theories can exhibit deconfined emergent electromagnetism at low energies. In a condensed matter context, apart from the quantum spin ice state of the pyrochlore XXZ model, a $U(1)$ liquid is expected also for the partially magnetized pyrochlore magnet with three of the four tetrahedral sublattice spins pointing along an applied field and the remaining spins anti-aligned along the field \cite{Bergman2006a,Bergman2006b}. The resulting spin model, in common with the XXZ model in zero field, maps to a dimer model which has been extensively studied \cite{Moessner2003,Bergman2006a,Bergman2006b,Sikora2009,Sikora2011}.  Apart from quantum magnets, there is theoretical and numerical evidence for the emergence of a gapless photon in models of exciton condensates \cite{SSLee1,SSLee2,SSLee3}, in a rotor model on a cubic lattice \cite{Motrunich2002,Motrunich2005} 
and there is a suggestion that protons in conventional water ice might be strongly correlated and, through quantum mechanical tunneling of the protons, also display a deconfined $U(1)$ phase \cite{CastroNeto2006}. There is also a proposed way of simulating a rotor model with emerging photons at low energies in a system of trapped cold atoms on a pyrochlore lattice \cite{Tewari}.

The quantum $U(1)$ liquids can be viewed as string-net condensate phases as made explicit in a three dimensional rotor 
model of Levin and Wen \cite{WenLevin2006}. String-net condensation is a framework within which a class of achiral gauge theories  can be understood \cite{WenLevin2006,WenLevin2005}. The idea, stated briefly, is that these phases can be obtained by specifying a lattice model with elementary bosonic degrees of freedom and a local Hamiltonian which condenses closed loops of different types into the ground state of a bosonic model. The Hamiltonian may also contain terms which give the condensed strings a tension which breaks the degeneracy of the resulting equal weight superposition of string states. Excitations of the string net condensate are the end points of closed strings and correspond to charges of some gauge theory. This framework offers a way to generalize the toric code of Kitaev \cite{Kitaev} by writing down an infinite set of exactly solvable models with Hamiltonians consisting of mutually commuting terms. 

Instabilities of the $U(1)$ liquid in ordered magnetic phases due to the condensation of visons (which are gapped in the deconfined phase) 
has been studied in several works \cite{Bergman2006b, Hermele2005, Alicea}. We also note that a three-dimensional 
$U(1)$ gauge theory coupled to a bosonic matter field has been found useful to describe transitions out of {\it classical} spin ice 
\cite{Powell2011}, through the condensation of the matter field, to give Higgs phases. In this context, the Higgs phases are magnetically ordered phases within the spin ice manifold.  


\section{A Materials Perspective}

\subsection{General considerations}
\label{sec:materials_considerations}

We discussed in Section \ref{sec:QSI} how the spin liquid state of 
quantum spin ice (QSI) materials originates from the (perturbative) anisotropic 
interactions between effective spins one-half away from the frustrated classical Ising limit.
Central to this story was the point that the model of Eqn.~(\ref{eq:QSI}) generates 
the crucial ring exchange Hamiltonian of Eqn.~(\ref{eq:ringexchange}) that
 drives classical spin ice into the $U(1)$ spin liquid state. Yet, we did not explain how Eqn.~(\ref{eq:QSI}) arises from the microscopic interactions between the magnetic ions. 
We explore in this subsection where the effective spin-1/2 model comes from and how it is 
amended when considering real quantum spin ice candidate pyrochlore materials such as the ones 
discussed in Subsection~\ref{sec:candidate_materials}.

In the search for QSI materials, we are {\it de facto} 
seeking systems with strong effective Ising-like anisotropy such that the order zero of the effective spin Hamiltonian 
is given by $H_{\rm CSI}$ in Eqn.~(\ref{eq:SI}). Such leading $H_{\rm CSI}$ interactions may arise from the single-ion anisotropy, as in (Ho,Dy,Tb)$_2$Ti$_2$O$_7$ or, 
it may be inherited from the interactions between the ions themselves, even though the single-ion anisotropy of 
a would-be isolated ion may {\it not} be Ising-like, as is the case for Yb$_2$Ti$_2$O$_7$ \cite{Ross2011,Applegate2012,Hayre2012}.
We are thus looking for materials that possess spin-orbit interactions that are stronger than the crystal field interaction 
and where the orbital angular momentum of the unpaired electrons is not quenched. In practice, one expects to find such 
a situation most prevalently among rare-earth (lanthanide, 4f  or actinide, 5f) elements. 
However, it is perhaps not ruled out that some materials based on 3d, 4d or 5d elements may eventually be found to display 
some of the classical or quantum spin ice phenomenology described above. 
For example, Co$^{2+}$ magnetic ions often exhibit a strong Ising-like anisotropy, as found in the quasi-one-dimensional 
Ising-like antiferromagnet CsCoBr$_3$ compound \cite{CsCoBr3}.  
In that context, one may note that the GeCo$_2$F$_4$ spinel, in which the magnetic Co$^{2+}$ ions reside on a (pyrochlore)
lattice of corner-sharing tetrahedra, has been reported to display a coexistence of spin ice, exchange and orbital frustration  \cite{GeCoF4}.
In the following, we restrict ourselves to insulating 
magnetic rare-earth pyrochlores of the form R$_2$M$_2$O$_7$ \cite{Gardner_RMP} where R$^{3+}$ is a trivalent 4f rare-earth 
element and M is a non-magnetic tetravalent ion such as Ti$^{4+}$, Sn$^{4+}$ or Zr$^{4+}$. 
Indeed, all materials for which QSI phenomenology has so far been invoked are based on rare earth pyrochlores.

The hierarchy of energy scales at play in rare-earth ions make them well suited for a material exploration of QSI physics.
In these systems, the spin-orbit interaction is larger than the crystal field energies, but not stronger than the intra-atomic 
electronic energy scale. Consequently, the spin-orbit interaction acts within the states defined by Hund's rules and
leads to a 2J+1 degenerate ionic ground state of spectroscopic notation $^{\rm 2S+1}$L$_{\rm J}$  where J=L+S if the 4f electron shell 
is more than half-filled and J=L-S otherwise. Note that S here is the electronic spin, not the pseudospin-1/2 ${\bm S}$ 
of the Hamiltonian in Eqn.~(\ref{eq:QSI}) and Eqn.~(\ref{eq:Heff}) below.
The effect of crystal-field perturbations (originating from the surrounding ligands) 
is to lift the degeneracy of the $^{\rm 2S+1}$L$_{\rm J}$ isolated ionic ground state. For the R$_2$M$_2$O$_7$ pyrochlores with 
Fd$\bar 3$m space group, odd-numbered electron (i.e. Kramers) ions (e.g. Dy, Er, Yb) have a magnetic ground state doublet 
and so do even-number electron (i.e. non-Kramers) 
(Pr, Tb and Ho) ions  \cite{Gardner_RMP}, but not Tm$^{3+}$ that has a non-magnetic singlet \cite{Zinkin_Tm2Ti2O7}.
In other words, the symmetry is sufficiently low for Kramers ions to cause them to have solely a doublet crystal field ground state, 
while the symmetry is still sufficiently high for non-Kramers ions for them (except Tm$^{3+}$) to have an accidental magnetic doublet ground state.

At this point, armed with the knowledge about the single-ion crystal field states  \cite{Bertin2012}, one can in, in principle, start to consider 
the inter-ionic interactions, $H_{\rm int}$,  and construct the pertinent microscopic (``UV'') Hamiltonian from which realistic 
amendments of Eqn.~(\ref{eq:QSI}) would result when perturbing the crystal field ground state with $H_{\rm int}$. 
It is here that the problem gets complicated -- especially when compared with, say 3d transition metal ion systems.
In these 3d systems, the angular momentum is typically quenched, the spin-orbit interaction is small 
and one is often dealing with a relatively simple spin-only exchange Hamiltonian of the form
$J_{ij} {\bm S}_i \cdot {\bm S}_j$. In system with 4f elements, except for Gd$^{3+}$,   
most of the angular momentum is provided by the angular momentum part of the atomic electrons and is not quenched. 
Furthermore, the relevant unfilled 4f orbitals are buried rather deep inside the inner part of the ion,  orbital overlap is reduced, and 
direct exchange or superexchange do not have the opportunity to completely dominate in $H_{\rm int}$.
Consequently, the ion-ion interactions among 4f systems end up having  a multitude of origins: 
direct classical electric and magnetic multipole interactions, electric and magnetic multipole interactions 
arising from direct exchange, superexchange electric and magnetic multipole interactions and lattice-mediated
 electric multipole interactions  \cite{4f5f_interactions}.
The microscopic couplings $\Omega_{ij}$, between ions $i$ and $j$ defining $H_{\rm int}$ are thus of a very high degree of complexity.
The microscopic Hamiltonian $H_{\rm int}$ can in principle be written in terms of Stevens equivalent operators
 $O({\mathbf J}_i)$. However, such a determination or parametrization, 
either experimentally or theoretically, of the pertinent (tensor) 
couplings $\Omega_{ij}$ between $O({\mathbf J}_i)$ and $O({\mathbf J}_j)$ 
(we have omitted here all angular momentum components that would define the various components of $\Omega_{ij}$) 
is enormously difficult to say the least.
 On the theoretical front, it is even questionable 
whether estimates of $\Omega_{ij}$ accurate to within 100\% of the true couplings could be achieved. 

Yet, things are not as hopeless as it appears.
For most cases of interest, we have a situation where the 
$\Omega_{ij}$ couplings are often of the order of 10$^{-2}$ K -- 10$^{-1}$ K or so, and are therefore typically
small compared to the gap $\Delta \sim 10^1-10^2$ K 
between the magnetic crystal field ground doublet and the first excited crystal field state(s). 
This means that the two pairs of crystal field doublet wavefunctions for two interacting ions $i$ and $j$ get very weakly 
admixed with the excited crystal field states via the action of $H_{\rm int}$. 
Out of the $(2J+1)^N$ crystal field states, where $N$ is the number of ions, one can consider only the 
subspace, $\Psi$, spanned by the $2^N$ individual crystal field ground states. 
One can then consider one of the many variants of degenerate perturbation 
theory to derive an effective $S=1/2$ ``pseudospin''  Hamiltonian,  
$H_{{\rm eff},\frac{1}{2}}$, describing the perturbed energies and eigenstates within $\Psi$.
Thanks to the relative high symmetry of the pyrochlore lattice, the projection of $H_{\rm int}$ into $\Psi$ gives for 
nearest-neighbours of interacting ions a relatively simple form for the effective nearest-neighbour interactions between
the pseudospins ${\bm S}_i$ and ${\bm S}_j$ for ions $i$ and $j$. These effective interactions, 
$H_{{\rm eff},\frac{1}{2}}$, are the ones that we seek to spell out the additional symmetry-allowed 
couplings in $H_{\rm QSI}$ in Eqn.~(\ref{eq:QSI}). 
Several notation conventions have appeared over the past two years 
for $H_{{\rm eff},\frac{1}{2}}$ \cite{Ross2011,thompson_PRL,mcclarty_JPCM,curnoe_PRB}.
 Here, we  adopt the one introduced in Ref.~[\onlinecite{Ross2011}] as it related most directly with Eqn.~\ref{eq:QSI}:
\begin{align}
& H_{{\rm eff},\frac{1}{2}} \\
	&=	\sum_{\langle i,j\rangle} \{ J_{\parallel}S_i^z S_j^z - J_\pm (S_i^+ S_j^- +S_i^-S_j^+)
       + J_{\pm\pm} [\gamma_{ij}S_i^+S_j^+  \nonumber \\ & +  \gamma_{ij}^* S_i^-S_j^-] 
      + J_{z\pm} [(S_i^z(\zeta_{ij}S_j^+ +\zeta_{i,j}^*S_j^-) + i \leftrightarrow j] \} .
\label{eq:Heff}
\end{align}
In Eqn.~(\ref{eq:Heff}), $\langle i,j \rangle $ refers to nearest-neighbour sites of the pyrochlore lattice,
$\gamma_{ij}$  is a $4 \times 4$ complex unimodular matrix, and $\zeta=-\gamma^*$ \cite{Ross2011,Savary_ETO}.
In this formulation, the effective ${\bm S}_i=1/2$ spins are now 
expressed in terms of a local triad of orthogonal unit vectors $\hat x_i$, $\hat y_i$ and
$\hat z_i$ with $\hat z_i$ along the local ``Ising'' $[111]$ direction, with $\pm$ referring to 
the two orthogonal complex directions $\hat x_i \pm i \hat y_i$.
As before, we have $H_{\rm CSI}= J_{\parallel} \sum_{\langle i,j\rangle} S_i^z S_j^z$, the classical (Ising) term responsible for the spin ice degeneracy 
at the nearest-neighbour level. The three other terms (proportional to $J_\pm$, $J_{\pm\pm}$ and $J_{z\pm}$) are {\it all} the extra
nearest-neighbour terms allowed by symmetry on the pyrochlore \cite{curnoe_PRB}, 
that do not commute with $H_{\rm CSI}$, and hence cause quantum dynamics within the classical spin ice manifold. 
One may ask what is the physical origin (or content) of the terms proportional to $J_{\pm\pm}$ and $J_{z\pm}$ in
$H_{{\rm eff},\frac{1}{2}}$. 
A simple perspective on this matter goes as follows.
One may consider the following four nearest-neighbour interactions on the pyrochlore lattice  \cite{mcclarty_JPCM}:
(i) an Ising interaction $J_\parallel S_i^z S_j^z$ as in Eqn.~(\ref{eq:Heff}),
(ii) an isotropic interaction of the form $J_{\rm iso} {\bm S}_i \cdot {\bm S}_j$, 
(iii) a pseudo-dipolar ``exchange'' interaction that has the same  ``trigonometric form'' as magnetostatic dipole-dipole, 
$J_{\rm pd} ( {\bm S}_i \cdot {\bm S}_j - 3 \left( {\bm S}_i \cdot \hat{\mathbf{r}}_{ij} \hat{\mathbf{r}}_{ij} \cdot {\bm S}_j \right)$
and, finally, 
(iv) a Dzyaloshinskii-Moriya interaction of the form $J_{\rm DM}( \hat{\mathbf{d}}_{ij} \cdot  {\bm S}_i \times {\bm S}_j)$ \cite{Canals2008}.
The set of interactions $(J_\parallel, J_{\rm iso}, J_{\rm pd}, J_{\rm DM})$ can be linearly transformed into the set 
$(J_\parallel, J_{\pm}, J_{\pm\pm}, J_{z\pm})$.

We are in the very early days of the systematic experimental and theoretical investigation of quantum spin ices. 
On the theoretical front, the immediate question is to determine what are the possible zero-temperature phases that 
$H_{{\rm eff},\frac{1}{2}}$ displays.
Using a form of gauge mean-field theory (gMFT), this question has been tackled for systems with 
 Kramers ions  \cite{Savary2011} as well as non-Kramers ions  \cite{Lee2012}  (non-Kramers ions must have
$J_{z\pm}=0$  \cite{Lee2012}). Recent work has extended this approach to investigate 
 the nonzero temperature phase diagram of this model  \cite{SavaryFT}. Examples of phase diagrams produced using this method are shown in Fig.~\ref{fig:gMFT}. The top two panels show phases expected through a section of the available space of nearest-neighbour couplings for odd electron magnetic ions. Of the two exotic phases in the top two panels of this figure, one is the quantum spin ice (QSI) phase which is the topic of this review. This appears, as we expect from perturbation theory in the vicinity of the classical spin ice point. The second (CFM) phase 
has co-existing symmetry broken long-range order, gapless photon excitations and gapped magnetic monopoles. The FM and AFM phases are respectively the six-fold degenerate ferromagnetic state with net moment along one of the the $\langle 001 \rangle $ directions and an antiferromagnetic phase with moments perpendicular to the local $ \langle 111 \rangle $ directions.
For non-Kramers ions, we refer to the bottom panel of Fig.~\ref{fig:gMFT} which shows the quantum spin ice and two quadrupolar phases.   One of the long-range ordered phases is ferroquadrupolar (FQ) and the other antiferroquadrupolar (AFQ).  The AFQ phase, in the language of effective pseudospin-1/2, is a coplanar antiferromagnetic phase while the FQ is the maximally polarized ferromagnet with moments perpendicular to the local Ising directions \cite{Lee2012}. 
 
For what concerns us here with this review, in terms of the existence of a $U(1)$ spin liquid, it probably suffices to say that as long as
$J_{\pm}$, either positive or negative, is the leading perturbation beyond $J_\parallel$, while being sufficiently larger than
$J_{\pm\pm}$ and $J_{z\pm}$, one finds a finite region over which the $U(1)$ spin liquid exists.
Magnetic rare-earth ions often possess a sizeable magnetic dipole moment. 
An obvious question is how the phase diagram of Refs.~[\onlinecite{Savary2011,Lee2012,SavaryFT}]
is modified by long-range magnetostatic dipole-dipole interactions.
Also, away from the $U(1)$ spin liquid phase of the phase 
diagram of Ref.~[\onlinecite{Savary2011,Lee2012}],  one may expect to encounter more complicated 
long-range ordered phases at nonzero ordering wavevector  \cite{Melko2001,SDSI,Javanparast}.

We end this subsection by a brief discussion  about how the coupling parameters 
$(J_\parallel, J_\pm, J_{\pm\pm},J_{z\pm})$ may be determined
experimentally -- a necessary task 
if one wants to rationalize the behaviour of real materials in relation to the theoretical phase diagrams \cite{Savary2011,Lee2012}.
First, we make a comment about how the complexity of the microscopic interactions 
$H_{\rm int} \sim \Omega_{ij} O({\mathbf J}_i)O({\mathbf J}_j)$ was swept aside when $H_{\rm int}$ was projected onto $\Psi$.
As long as $\vert \Omega_{ij}\vert \ll \Delta$, one can calculate any correlation functions involving the observable 
${\mathbf J}_i$ angular momentum operators via calculation of 
correlation functions involving the effective ${\bm S}_i$ spins. In such a case, the 
theory is quantitative and characterized by the four 
$(J_\parallel, J_\pm, J_{\pm\pm},J_{z\pm})$ couplings (to be determined by experiments) along with the magnetic dipole-dipole interaction and the matrix elements of ${\mathbf J}_i$ within the non-interacting ground state doublet (involving the so-called single-ion $g$-tensor)  \cite{Bertin2012}.
 However, if $H_{\rm int}$ significantly admixes excited crystal field states into the ground state doublet, 
observables involving ${\mathbf J}_i$ are no longer trivially related to the pseudospins $\mathbf{S}_i$, 
and the theory is no longer quantitative if the microscopic $\Omega_{ij}$ couplings are not known.
We return to this in the Subsection \ref{sec:Tb2Ti2O7} when discussing the Tb$_2$(Ti,Sn)$_2$O$_7$ compounds.
The situation is conceptually similar to the problem of calculating the staggered magnetization, $M^\dagger$, of the simple square lattice
one-band Hubbard model at half-filling when recast as an effective spin-1/2 Hamiltonian. Away from the Heisenberg limit, in which
the hopping $t$ is infinitely small compared to the Hubbard $U$ (i.e. $t/U \ll 1$), 
$M^\dagger$ is a function of $t/U$, and is no longer determined by the textbook formula of the 
thermodynamic average of the staggered $z$ component of localized spin operators, $S_i^z$, given by
$M^\dagger = \sum_i (-1)^i \langle S_i^z\rangle$  \cite{Delannoy_PRB-1}. The problem becomes
even more complicated when the microscopic model is that of an extended Hubbard model  \cite{Delannoy_PRB-2}.
In order to calculate the staggered magnetization with the effective spin-1/2 model the corresponding
operator must be defined in the microscopic (electronic) Hubbard theory first, and then canonically transformed in the 
effective low-energy (spin-1/2) theory  \cite{Delannoy_PRB-1,Delannoy_PRB-2}.

Following the initial realization that anisotropic exchange may be of importance in R$_2$M$_2$O$_7$ materials  \cite{curnoe_PRB}, a number of experimental
studies, mostly using elastic and inelastic neutron scattering,  have been targetted to determine the value of these couplings.
These studies fall in two categories. A first category has assumed that the nearest-neighbour part of the interactions 
$H_{\rm int}$ between ${\mathbf J}_i$ angular momentum operators are bilinear and of anisotropic nature  \cite{thompson_PRL,cao_PRL,cao_JPCM,malkin,thompson_JPCM}. 
The other group consists of studies that do
not make this assumption \cite{Ross2011,Savary_ETO,chang} but work, instead,  with a model with anisotropic exchange between 
pseudospin-1/2 degrees of freedom as in Eqn.~(\ref{eq:Heff}).
 Ultimately, from the discussion above regarding the need to transform observables, one concludes that if the crystal field
gap $\Delta$ is not very large compare to the $\Omega_{ij}$ interactions in $H_{\rm int}$, the theory and the ultimate description of the data is not on 
on a very strong footing anyway from the word go. 
This is particularly the case for Tb$_2$Ti$_2$O$_7$  \cite{Molavian2007,Molavian2009}. In the other extreme, for example in Yb$_2$Ti$_2$O$_7$, 
the gap $\Delta \sim 600$ K \cite{Bertin2012} 
is so large that one can employ either a pseudospin-1/2 representation  \cite{Ross2011,chang}, or a model with bilinear couplings between the
${\mathbf J}_i$  \cite{thompson_PRL,thompson_JPCM}. 
In that case the anisotropic bilinear 
exchange couplings, $\Omega_{ij}$, between ${\mathbf J}_i$ and ${\mathbf J}_j$ are more or less an exact inverse linear transformation of the couplings 
between the ${\bm S}_i$ pseudospins \cite{thompson_PRL}.
 The same argument probably applies to Er$_2$Ti$_2$O$_7$ where a spin-1/2 model  \cite{Savary_ETO} and one
with ${\mathbf J}_i - {\mathbf J}_j$ couplings  \cite{mcclarty_JPCM,cao_JPCM} can be used for zero field of for magnetic fields less than a few tesla, 
though we note that this material is no quantum spin ice candidate. In the case of Yb$_2$Ti$_2$O$_7$, a recent paper  \cite{Hayre2012} has discussed the pitfalls of
some of the analysis and models employed in a number of works \cite{thompson_PRL,cao_PRL,cao_JPCM,malkin}.

Staying with systems that can be described by an effective spin-1/2 model as in Eqn.~(\ref{eq:Heff}), two neutron scattering methods have been used to determine the $\{J_e\} \equiv (J_\parallel, J_\pm, J_{\pm\pm},J_{z\pm})$ couplings. One method is to fit the energy integrated paramagnetic scattering, $S({\bm q})$, obtained at sufficiently high temperature, and compare it
with the corresponding $S({\bm q})$ obtained via mean-field theory \cite{enjalran_PRB,kao_rpa}. If wanting to neglect the long-range dipolar interactions, one can also fit the experimental $S({\bm q})$ at sufficiently high temperature with that calculated on the basis of an approximation using finite clusters \cite{dalmas_eto}. Such an approach could be formally rationalised on the basis of the so-called numerical linked cluster method (NLC) \cite{Applegate2012,Hayre2012}. It is probably fair to say that mean-field methods have not yet
successfully yielded accurate values for the $\{J_e\}$ couplings for any of the R$_2$Ti$_2$O$_7$ materials considered. The main technical difficulty is that one is required to perform experiments in a temperature regime which is at least 5 to 10 times higher than the mean-field transition temperature
($T_c^{\rm mf}$) of the underlying spin-1/2 Hamiltonian in order for these methods to be quantitative. At such high relative temperatures, the intensity modulation of the experimental $S({\bm q})$ can be quite weak and difficult to fit.
Previous works on Yb$_2$Ti$_2$O$_7$  \cite{thompson_PRL,chang} have considered temperature that are actually below $T_c^{\rm mf}$ and the reported values of $\{J_e \}$ couplings are, 
consequentially, quite inaccurate \cite{Hayre2012}. 
A work on Er$_2$Ti$_2$O$_7$ that compares the experimental $S({\bm q})$, 
obtained at a temperature reasonably high compared to the true critical temperature, 
with that obtained from calculations on a single tetrahedron, may be less subject to this concern \cite{dalmas_eto}.

The other method that is currently enjoying some popularity employs inelastic neutron scattering to probe the excitations in the 
field-polarized paramagnetic state \cite{Ross2011,Savary_ETO}. 
By fitting the dispersion and intensity of these excitations compared to those calculated from the model (\ref{eq:Heff}), 
supplemented by a magnetic Zeeman field term, one can obtained the four couplings $\{J_e\}$.
This approach has been applied to determine (fit) 
the $\{J_e\}$ for both Yb$_2$Ti$_2$O$_7$ \cite{Ross2011} and Er$_2$Ti$_2$O$_7$ \cite{Savary_ETO}.
In these fits, the magnetostatic dipole-dipole interactions have only been considered at the nearest-neighbour level, 
and their nonzero value are thus implicitly folded in the fitted $\{J_e\}$ values.
There is nothing in principle that would prevent performing an analysis of these in-field excitations 
that would include the true long-range dipolar interactions \cite{maestro_JPCM,maestro_PRB}
 and thus determine the ``real'' $\{ J_e \}$ nearest-neighbour values. 
In view of the fact that the current $H_{\rm eff}$ model in Eq.~(\ref{eq:Heff}) neglects effective
exchange couplings between the pseudospin-1/2 beyond nearest-neighbours, which have been found
in the classical dipolar spin ice Dy$_2$Ti$_2$O$_7$ to be about 10\% of the nearest-neighbour $J_\parallel$  \cite{Yavorskii},
it is indeed perhaps justifiable to ignore dipolar interactions beyond nearest-neighbours altogether.
In that context, it is interesting to note that recent calculations 
that make use of a sort of series expansion method based on the numerical linked-cluster (NLC)
expansion, has shown that the specific heat  \cite{Applegate2012} and the magnetic field and temperature 
dependence of the magnetization  \cite{Hayre2012} of Yb$_2$Ti$_2$O$_7$ is well described using 
the $\{J_e\}$ determined by the aforementioned fit to inelastic neutron scattering data \cite{Ross2011}.

\subsection{Candidate materials}
\label{sec:candidate_materials}

In this section we briefly discuss four materials among the R$_2$Ti$_2$O$_7$  familly that may be 
candidates for displaying some of the quantum spin ice (QSI) phenomenology.
These are Tb$_2$Ti$_2$O$_7$, Pr$_2$M$_2$O$_7$ (M=Sn,Zr) and Yb$_2$Ti$_2$O$_7$.
While these compounds display a number of attributes that warrant discussing their 
exotic thermodynamic properties in the context of QSI physics,
it is fair to say that, at the time of writing,  
that there is no definitive evidence that any one of them displays a quantum spin ice state.

\subsubsection{Tb$_2$Ti$_2$O$_7$}
\label{sec:Tb2Ti2O7}

This was the first material for which the name {\it quantum spin ice} was coined
 \cite{Molavian2007}. Upon cooling, Tb$_2$Ti$_2$O$_7$ 
starts to develop magnetic correlations at a temperature of 20 K or so, 
but most experimental studies have so far failed to observe 
long-range order down to the lowest temperature \cite{Gardner1999,Gardner2003}.
Some early reports found signs for slow dynamics below $\sim 300$ mK  \cite{Gardner2003}.
suggesting some kind of spin-freezing/spin-glassy  phenomena  \cite{Luo2001}. 
With overall antiferromagnetic interactions, 
indicated by a negative Curie-Weiss temperature $\theta_{\rm CW}$,
one would naively expect this non-Kramers Ising system to display 
a non-frustrated long-range ordered ground state 
with  all ``up'' tetrahedra in Fig.~\ref{fig:pyrochlore}
having their four spins pointing  ``in'' and all ``down'' tetrahedra having spins pointing ``out'',
or vice-versa \cite{Harris_JPCM,Moessner_FM_PRB}. 
By considering an Ising dipolar spin ice model with competing nearest-neighbour antiferromagnetic  ($J_\parallel < 0$ in Eqn.~(\ref{eq:SI})) and long-range dipolar interactions,
Ref.~[\onlinecite{denHertog}] found a critical temperature around 1 K, 
in total disagreement with experiments \cite{Gardner1999}.
Further evidence that such an Ising model was too simple for Tb$_2$Ti$_2$O$_7$ came 
from neutron scattering experiments \cite{Gardner_Waldron}.
These found a broad region in reciprocal space near the point ${\bm q}=002$ 
with high scattering intensity  \cite{Gardner2003,Gardner_Waldron}  which is inconsistent with
what is naively expected for an Ising model \cite{Enjalran_PRB}. 
Subsequent theoretical work  \cite{kao_rpa,Enjalran_PRB} found that allowing for the magnetic moments to 
 fluctuate transverse to their
local $[111]$ Ising direction could lead to a high scattering intensity at ${\bm q}=002$.
These early observations, along with the fact that such broad ${\bm q}=002$ intensity 
remains down to a temperature of 50 mK  \cite{Gardner2003}
made it clear that one or more mechanisms had to be considered to generate non-Ising fluctuations and response down to the lowest temperature.

The classical spin ice compounds Ho$_2$Ti$_2$O$_7$ and Dy$_2$Ti$_2$O$_7$ have a large gap $\Delta$ of order of 300 K 
 between their crystal field ground state doublet and their first excited doublet. As discussed in Section \ref{sec:materials_considerations}, 
this feature is at the origin of an Ising description of these systems. 
In contrast, Tb$_2$Ti$_2$O$_7$ has $\Delta \sim 18$ K \cite{Gingras2000}.
Calculations of the dynamical structure factor $S({\bm q},\omega)$ in the paramagnetic phase
that employs the random phase approximation (RPA) method  \cite{kao_rpa}
make a strong point that this small gap  $\Delta$ allows for a significant admixing 
between the crystal field states that is induced by the interactions among the ${\mathbf J}_i$ 
angular momenta through superexchange and long-range dipolar interaction.  These effects are 
rendered even more significant since it appears that, at the level of an Ising model description 
that ignores excited crystal field states \cite{denHertog}, Tb$_2$Ti$_2$O$_7$ is near a boundary between an
``all-in/all-out'' long range ordered phase and a ``2-in''/``2-out'' spin ice state  \cite{denHertog}. In other words,
 since projected interactions $H_{\rm int}$ in the crystal field ground state puts the material near a
 phase boundary between distinct classical ground states, corrections beyond this projection that involve the details of the 
${\mathbf J}_i-{\mathbf J}_j$ interactions, $\Omega_{ij}$,  and the excited crystal field states must be revisited 
and incorporated into the effective low-energy Hamiltonian.

Considering a simple model of nearest-neighbor isotropic exchange $\mathbf{J}_i \cdot \mathbf{J}_j$ between the ${\mathbf J}_i$ 
angular momenta as well as long-range dipole-dipole interactions, 
Refs.~[\onlinecite{Molavian2007,Molavian2009}] found, via second order perturbation theory calculations, 
 that the original dipolar Ising spin ice model description of Tb$_2$Ti$_2$O$_7$ is significantly 
modified. Rather, an early form of an  
effective low-energy pseudospin-1/2 model similar to that of  Eqn.~(\ref{eq:Heff}), 
supplemented by further anisotropic exchange terms as well as long-range dipolar interactions, was obtained  \cite{Molavian2009}.
One may thus consider that Tb$_2$Ti$_2$O$_7$ is one system for which a quantum spin ice Hamiltonian of 
the form such as in Eqn.~(\ref{eq:Heff}) is well motivated. The lack of long-range order in Tb$_2$Ti$_2$O$_7$ is thus, 
perhaps, intriguingly related to the $U(1)$ spin liquid of Eqn.~(\ref{eq:Heff}) \cite{Molavian2007}.

Over the past couple of years, however, the situation regarding the phase of 
Tb$_2$Ti$_2$O$_7$ below a temperature of $2$ K has become perhaps ever more cloudy.
The old evidence \cite{Mamsurova} for a sizeable magneto-elastic response in this system has 
resurfaced \cite{Mirebeau_Nature,Ruff2007,Malkin_Tb2Ti2O7,Fennell2013}.
 Some authors have proposed that, possibly related to this lattice effect, the crystal field ground state of Tb$^{3+}$ 
is split into two singlets separated by an energy gap of order of 2 K, and that this is the principal reason why this system does not 
order  \cite{Bonville_singlet,Petit_singlet}. 
Interestingly, the closely related Tb$_2$Sn$_2$O$_7$ compound develops long-range order at 0.87 K into a ${\bm q}=0$ 
long-range ordered version of a ``2-in''/``2-out'' spin ice state, but with the magnetic moments canted away from the strict $\langle 111 \rangle$ Ising 
directions  \cite{Mirebeau_Tb2Sn2O7}. 
Tb$_2$Sn$_2$O$_7$ has also been proposed to have such a split crystal field ground doublet  \cite{Petit_Tb2Sn2O7}. 
But, with the suggestion that it has stronger interactions than its Tb$_2$Ti$_2$O$_7$ cousin, 
Tb$_2$Sn$_2$O$_7$ is argued to overcome the formation of a trivial
non-magnetic singlet ground state and to develop long-range order.
The suggestion that there exist an inhibiting doublet spliting mechanism leading to a 
singlet-singlet gap as large as 2 K  in Tb$_2$Ti$_2$O$_7$ is currently being debated \cite{Gaulin_no_gap}.

Some recent neutron scattering work on Tb$_2$Ti$_2$O$_7$ finds evidence for pinch-points suggesting
 the presence of ``2-in''/``2-out'' spin ice -like correlations (see discussion in Section ~\ref{sec:spin_ice}) \cite{Fennell2012}.
Even more recent inelastic neutron scattering studies have identified some 
clear and reasonably intense features at the ${\bm q} = \frac{1}{2} \frac{1}{2} \frac{1}{2}$  
reciprocal space point lattice point suggesting the development of nontrivial magnetic correlations
at the lowest temperature and which have been referred to as ``antiferromagnetic spin ice correlations'' \cite{Petit_singlet,Fritsch,Guitteny2013,Kadowaki_Tb2Ti2O7}.
 This, along with the observation of elastic magnetic 
scattering below an energy of 0.05 meV $\sim$ 0.5 K,  may be viewed as inconsistent with
the above non-magnetic singlet ground state scenario \cite{Bonville_singlet,Petit_singlet}. 
Finally, it has become clear over the past 
two years that there are significant 
sample-to-sample variations in the thermodynamic properties exhibited among single crystals of 
Tb$_2$Ti$_2$O$_7$  \cite{Gaulin_no_gap,Kadowaki_Tb2Ti2O7,Chapuis_Tb2Ti2O7}. 
This may ultimately be the cherry on the cake in terms of the plethora of phenomena 
Tb$_2$Ti$_2$O$_7$ displays.
The sample-to-sample variability may be endorsing a picture that this compound is naturally located 
near the vicinity of a transition between two (or more) competing states.
There is probably sufficient evidence in place suggesting that there are spin-ice 
like correlations and transverse fluctuations of the angular momenta in Tb$_2$Ti$_2$O$_7$ 
so that a quantum spin ice picture is not, at this time, ruled out. 
However, there is definitely more to the story and there are numerous 
 hints that the lattice degrees of freedom are not inert bystanders in
Tb$_2$Ti$_2$O$_7$, and that magneto-elastic couplings should probably be considered carefully.
In that context, the possibility of magneto-elastic interactions and the concurrent existence 
of symmetry related quadrupolar-like interactions naturally brings up the question:
``what is the role of  quadrupolar-like interactions in even electron (non-Kramers) magnetic ion
 systems in modifying the simplest (Ising magnet) description of these systems?'' 
This question has been explored with the quantum spin ice candidates, Pr$_2$M$_2$O$_7$ (M=Sn,Zr), that we next discuss.

\subsubsection{Pr$_2$Sn$_2$O$_7$ \& Pr$_2$Zr$_2$O$_7$}
\label{sec:Pr2M2O7}

It might be said that the modern research era in frustrated quantum spin systems was, at least partially, 
triggered by Anderson's 1987 Science paper  \cite{Anderson}. In this paper, Anderson noted 
that geometrical frustration might naturally lead to exotic quantum states of magnetic matter, including resonating
valence bond (RVB) states from which unconventional superconductivity may arise.
Unfortunately, if the field of condensed matter physics has long been experiencing
a drought in the abundance of (quantum) spin liquid candidate materials \cite{PALee_drought}, 
the scarcity of highly frustrated magnetic materials that are 
at the verge of a Mott-insulator transition, or display simultaneously frustrated localized magnetic moments along with
itinerant electrons, may remind one of the Martian atmosphere.
From that perspective, frustration and development of superconductivity in organic Mott insulators is more than a curiosity \cite{PowellMcKenzie}.
In that context, it is perhaps not surprising that 
the discovery of spin-ice like ``2-in''/``2-out'' correlations, 
signalling geometrical frustration for would-be isolated non-Kramers Pr$^{3+}$ ions,
in the {\it metallic} Pr$_2$Ir$_2$O$_7$ pyrochlore compound  \cite{Nakatsuji2006,Machida2010}
 has attracted a fair amount of interest  \cite{Flint2012,Rau2013,LeeSungBin2013}.
Before attacking the complexities of Pr$_2$Ir$_2$O$_7$, such as the resistivity minimum as a function of temperature
(Kondo-like effect) \cite{Nakatsuji2006}  and anomalous Hall effect \cite{Machida2010}, one may wonder whether, even in insulating Pr-based
compounds, there might exist unusual properties that may be of relevance to the physics of metallic Pr$_2$Ir$_2$O$_7$.

The pyrochlore form of Pr$_2$Ti$_2$O$_7$ does not exist \cite{Gardner_RMP} at ambient temperature and pressure. 
However, the insulating and magnetic Pr$_2$Sn$_2$O$_7$ and Pr$_2$Zr$_2$O$_7$ compounds do exist and both form a regular pyrochlore structure.
In these two materials, the Pr$^{3+}$ non-Kramers ions possess a magnetic crystal field  Ising doublet ground state, as 
the Ho$_2$Ti$_2$O$_7$ spin ice and the above paradoxical Tb$_2$Ti$_2$O$_7$.
Pr$_2$Sn$_2$O$_7$ is, unfortunately, not amenable to single-crystal growth using modern image furnace methods, but Pr$_2$Zr$_2$O$_7$ has been grown
successfully. Interestingly, neither material appears to develop long-range order down to temperatures 
of the order of $50$ mK. AC susceptibility measurements found a spin freezing in Pr$_2$Sn$_2$O$_7$ below a temperature of order of $100-200$ mK  \cite{Matsuhira2004}.
Powder neutron diffraction on Pr$_2$Sn$_2$O$_7$ reveal short range correlations and, surprisingly,
it has a residual low-temperature entropy larger than the Pauling entropy $S_0$ found in classical spin ices \cite{Zhou2008}.
Inelastic neutron scattering suggests that the low-temperature state of this system remains dynamic down to at least 200 mK 
 \cite{Zhou2008}.

Early work on Pr$_2$Zr$_2$O$_7$  \cite{Matsuhira2009} 
found a negative Curie-Weiss temperature of $-0.55$ K  (determined below a temperature of 10 K), indicating 
effective antiferromagnetic interactions between the magnetic moments.
AC magnetic susceptibility measurements did not find evidence for a transition to 
long-range order down to $80$ mK. However, some frequency dependence
of the AC susceptibility was observed below $0.3$ K, indicating some form of spin freezing.
More recent work on single crystals of  Pr$_2$Zr$_2$O$_7$
report evidence of spin ice -like correlations and quantum fluctuations  \cite{Wen_Pr2Zr2O7}.
 Heat capacity and magnetic susceptibility measurements show no sign of long-range order down to $50$ mK. The wave vector dependence of quasi-elastic neutron scattering at $100$ mK
shares some similarities with one of a classical Ising spin ice, including pinch-points. These results are interpreted as an indication of
``2-in''/``2-out'' ice rule being satisfied over a time scale set by the instrumental energy resolution.
Quite interesting, and in sharp contrast with classical spin ices where almost no inelastic response is observed \cite{Clancy}, 
inelastic scattering in Pr$_2$Zr$_2$O$_7$ with an energy transfer of $0.25$ meV does not show pinch points. 
This suggests that there are fluctuations operating which break the ice rule.

In summary, it appears that the insulating Pr$_2$(Sn,Zr)$_2$O$_7$ compounds develop significant correlations 
at temperatures below approximately $1$ K, but do not develop true long range order 
nor do they appear to behave like the conventional 
Ho$_2$Ti$_2$O$_7$ and Dy$_2$Ti$_2$O$_7$ classical dipolar spin ices.
As in these latter systems, the lowest excited crystal field levels in Pr$_2$(Sn,Zr)$_2$O$_7$ are at high energy, with a gap $\Delta \sim O(10^2)$  K 
above the ground doublet  \cite{Zhou2008}.
Consequently, the virtual crystal field fluctuation mechanism induced by the interactions between the magnetic moments,
and proposed to be at play in Tb$_2$Ti$_2$O$_7$ because of its small gap $\Delta ~18$ K  \cite{Molavian2007,Molavian2009},
is likely not significant in these two Pr-based compounds. It has been suggested that the microscopic $H_{\rm int}$ interactions 
in these Pr-based compounds contain strong multipolar interactions between the ${\mathbf J}_i$ 
angular momenta operators  \cite{OnodaTanaka2011,OnodaTanaka2010} and that these introduce quantum fluctuations 
that ``melt'' the low temperature classical spin ice state that would have developed in their absence.
The idea, although not quite presented in this form in 
Refs.~[\onlinecite{OnodaTanaka2011,OnodaTanaka2010}] is that high multipolar interactions
between the ${\mathbf J}_i$ operators, which involve higher powers of ${\mathbf J}_i$ than simpler bilinear exchange-like couplings $\Omega_{ij}^{uv} J_i^u J_j^v$, where $u,v$ are
cartesian components, have large matrix elements between the two states forming the crystal field doublet ground state of
Pr$^{3+}$ in Pr$_2$(Sn,Zr)$_2$O$_7$. At the end of the day, 
the projection of the derived complex microscopic inter-ionic Pr-Pr interactions onto the crystal field ground doublet  
leads to transition matrix elements between the two states, $\vert \psi^{\pm} \rangle$ that make up the doublet, these
 bringing about quantum dynamics between $\vert \psi^{+} \rangle$ and $\vert \psi^{-} \rangle$. 
An effective pseudospin-1/2 Hamiltonian, which describes that physics, 
can then be constructed and found to be of the form of $H_{{\rm eff},\frac{1}{2}}$, but with $J_{z\pm}=0$ since degenerate states of non-Kramers ions have vanishing 
matrix elements of time-odd operators such as ${\mathbf J}$. 
We thus reach the interesting conclusion that Pr-based quantum spin ice candidates are rather attractive from the perspective of systematic experimental {\it and} theoretical investigations:
(i) the effect of the excited crystal field states can probably be safely ignored, 
(ii) the magnetic moment $\mu\sim 3$ $\mu_{\rm B}$ means that dipolar interactions are 10 times weaker than in the
10 $\mu_B$  Ho$_2$Ti$_2$O$_7$ and Dy$_2$Ti$_2$O$_7$ classical spin ices materials \cite{denHertog,Siddharthan} and can be neglected as a first approximation down 
to about 0.1 K.
and, being non-Kramers ions, their effective pseudospin-1/2 Hamiltonian consists of only three couplings 
($J_\parallel$, $J_\pm$, $J_{\pm\pm}$), making the theoretical description of these materials more sober in terms of number of free parameters.

The systematic theoretical \cite{Lee2012,SavaryFT,OnodaTanaka2011,OnodaTanaka2010} and experimental 
exploration of quantum spin ice physics in Pr-based materials has now begun in earnest \cite{Zhou2008,Wen_Pr2Zr2O7} 
and one may expect exciting results to emerge from future studies.
That said, given the difficulty involved in synthesizing clean Sn-based and Zr-based pyrochlore oxides  \cite{Gardner_RMP}, one could, 
or perhaps even will, always worry about the effect that weak/dilute 
non-symmetry-invariant perturbations may have on the Pr-based materials 
given that the magnetic crystal field ground state is not  protected by the Kramers theorem.
The above discussion about Tb$_2$Ti$_2$O$_7$ and Pr$_2$(Sn,Zr)$_2$O$_7$ allow us to rationalize the naturalness of the next class of materials candidates for the study of quantum spin ice phenomenology.
We desire materials with large energy gaps $\Delta$ between their crystal field ground state and their first excited energy level.
Ideally, they should have small magnetic dipole moments (say less than 3 $\mu_{\rm B}$) so that dipolar interactions may be neglected, at least initially. Were it not for the concern of disorder breaking their crystal field degeneracy, non-Kramers ions would seem appealing candidates because the single ion dipolar doublets are Ising-like. The crystal field doublets of Kramers ions are stable against small local crystal field deformations that may be caused by imperfect sample quality but the low energy anisotropy is not constrained to be Ising-like and, indeed, rare earth Kramers ions are known that span the range from Ising to Heisenberg to XY spins. While the extent of the available parameter space might disfavour Kramers magnets as potential quantum spin ice candidates, it certainly does not rule them out and in fact, Yb$_2$Ti$_2$O$_7$, which we discuss next, is attracting much current interest in this respect.

\subsubsection{Yb$_2$Ti$_2$O$_7$}
\label{sec:Yb2Ti2O7}

Going back to some of the very earliest experimental studies of magnetic pyrochlore oxides,
 Bl\"ote and co-workers had observed in Yb$_2$Ti$_2$O$_7$  a broad specific heat bump at a temperature 
of about $2$ K, followed at lower temperature by a sharp specific heat peak at a critical temperature of $T_c \sim 0.214$ K 
suggesting a transition to long-range order \cite{Blote1969}. 
It was not until until the late 1990s and early 2000s that this compound was reinvestigated  \cite{Siddharthan,Hodges2002}, with the previously observed  \cite{Blote1969} 
sharp specific heat transition confirmed. The work of Ref.~[\onlinecite{Hodges2002}] reported data from
$^{170}$Yb M\"ossbauer and muon spin relaxation (muSR) 
measurements revealing a rapid collapse of the spin fluctuation rate just above $T_c$. 
However, powder neutron diffraction did not find signs of long-range
magnetic order below $T_c$ and muSR found a temperature-independent muon spin depolarization rate below $T_c$ which  was interpreted as a quantum fluctuation regime.
The work of Ref.~[\onlinecite{Hodges2002}]  provided strong evidence that the Yb$^{3+}$ ion 
in Yb$_2$Ti$_2$O$_7$ should be viewed as an XY system (i.e. with $g$ tensor components 
$g_\perp > g_\parallel$), meaning that the  
magnetic moments have their largest magnetic response perpendicular to the local $[111]$ direction. 
A subsequent single-crystal neutron scattering study
reported evidence for ferrimagnetic order \cite{Yasui2002}, but this was
soon contested by neutron depolarization measurements  \cite{Gardner_YbTO_2004}.
In Refs.~[\onlinecite{Bonville2004}] and \cite{Ross2009},
neutron scattering measurements revealed the development of rods of scattering 
intensity along the $\langle 111 \rangle$ directions, 
which was interpreted as the presence of quasi-two-dimensional spin correlations, an interesting and unusual 
phenomenon, assuming this interpretation to be correct, for  a three-dimensional cubic system.
These rods of scattering were found to be present at a temperature as high as $1.4$ K \cite{thompson_PRL}.
The application of a magnetic field as low as $0.5$ Tesla along the $[110]$ direction was found to induce
a polarized three-dimensional order accompanied by spin waves  \cite{Ross2009}.
Polarized neutron scattering measurements found, through an analysis of the neutron spin-flip ratio \cite{cao_PRL,cao_JPCM},
a non-monotonic temperature evolution of the component of the local spin susceptibility, $\chi_{\rm loc}$, 
parallel to the local $[111]$ direction.
Such behaviour is surprising for an XY system with $g_\perp > g_\parallel$, and
the behavior observed for $\chi_{\rm loc}$ provided compelling evidence for strongly anisotropic 
effective exchanges at play in Yb$_2$Ti$_2$O$_7$, with a very strong effective
Ising  exchange $J_\parallel$ term as in Eqn.~(\ref{eq:Heff}).

As discussed in Section \ref{sec:materials_considerations}, 
several works \cite{Ross2011,thompson_PRL,cao_PRL,cao_JPCM,malkin,chang} 
have endeavoured to determine the strength of the interactions in Yb$_2$Ti$_2$O$_7$.
It appears that the effective coupling between pseudospins $1/2$ are in fact strongly anisotropic, 
with the largest one being indeed $J_\parallel$. Among all values having been reported, it appears that
the exchange parameters determined in Ref.~[\onlinecite{Ross2011}]
 describe the bulk thermodynamic properties of the material reasonably well,
 at least down to $0.7$K \cite{Applegate2012,Hayre2012}.
Quite recently, the previously debated \cite{Gardner_YbTO_2004}
 report of a long-range ferrimagnetic order \cite{Yasui2002} in Yb$_2$Ti$_2$O$_7$ has been
 reconfirmed  \cite{chang}.  
In this ferrimagnetic state, the magnetic moment are found to be predominantly aligned along one of the
six $\langle 100\rangle$ cubic directions, but slightly splayed away from complete alignment, 
hence the label ferrimagnetic state.
In this context, it is worth mentioning a very recent paper 
on the closely related Yb$_2$Sn$_2$O$_7$ material \cite{Yaouanc2013}.
In this latter work, specific heat, $^{170}$Yb M\"ossbauer, 
neutron diffraction and muon spin relaxation measurements 
on powder samples find a first order transition at $0.15$ K to a state  
that M\"ossbauer and  neutron diffraction suggest to be the above ferrimagnetic order, referred to as
long-range ``splayed ferromagnetic'' order.
The situation about the nature of the low-temperature state of Yb$_2$Ti$_2$O$_7$ is thus rather confusing.
It may be  potentially illuminating to know that there is significant 
sample-to-sample variability among single crystal samples as inferred from the sharpness of the $T_c \sim 0.2$ K specific heat peak 
\cite{DOrtenzio2013,yaouanc_YbTO_sample,ross_YbTO_sample,ross_YbTO_stuff}. 
A recent extensive structural study investigation \cite{ross_YbTO_stuff}
has identified at least one origin of these variations:
single crystals grown via the floating zone technique show, compared to sintered powder samples, that up to
2.3 \% of the non-magnetic Ti$^{4+}$ sites get replaced by magnetic by Yb$^{3+}$.
Such ``stuffing'' of the transition metal ion site Yb$^{3+}$ ions  would introduce random exchange bonds and local
 lattice deformations and these may be at the origin of the mechanism
affecting the stability of the magnetic ground state of a would-be structurally perfect Yb$_2$Ti$_2$O$_7$ material. 
Finally, we note that new and very recent muon spin relaxation results find no evidence for the development of static order in either powder or single crystal samples of Yb$_2$Ti$_2$O$_7$ and, unlike in Ref~[\onlinecite{Hodges2002}] find no rapid collapse of the Yb$^{3+}$ spin fluctuation rate upon approaching the transition at $\sim 0.2$ K from above~\cite{DOrtenzio2013}.

The overall situation with Yb$_2$Ti$_2$O$_7$ is thus as follows. 
The effective exchange interactions are strongly anisotropic. 
On the basis of the determined \cite{Ross2011} and 
reasonably well validated \cite{Applegate2012,Hayre2012}  exchange  couplings, 
simple mean-field theory \cite{Ross2011}, classical ground state energy minimisation \cite{Wong2013}, and more sophisticated gauge 
mean-field theory calculations \cite{Savary2011,SavaryFT} predict a conventional long-range ferrimagnetic order
with the spins slightly splayed away from from the six cubic $\langle 111 \rangle$ directions as recently reported for Yb$_{2}$Sn$_{2}$O$_{7}$ \cite{Yaouanc2013} and characterized
by negligible quantum fluctuations \cite{Ross2011}. 
On the basis of these same calculations, Yb$_2$Ti$_2$O$_7$ is predicted to be located deeply in this semi-classical 
splayed ferromagnetic state, away from any phase transition boundaries with other conventional 
classical long-range ordered phases \cite{Wong2013}, or with the $U(1)$  quantum spin liquid phase or yet with 
the unconventional quantum Coulomb ferromagnetic (CFM) phase that Refs.~[\onlinecite{Savary2011,SavaryFT}] predict.
It therefore seems likely that neither Yb$_2$Ti$_2$O$_7$ nor Yb$_2$Sn$_2$O$_7$ \cite{Yaouanc2013} are 
good realizations of the sought $U(1)$ liquid in a quantum spin ice setting.
That said,  with a tendency towards a broken discrete symmetry state (i.e. ferrimagnetic 
order along one of $\langle 100\rangle$ directions) characterized by gapped excitations throughout the Brillouin zone, 
it is rather unclear why this material should be so sensitive to
dilute disorder such as the one generated by Yb$^{3+}$ stuffing on Ti$^{4+}$ sites \cite{ross_YbTO_stuff}.
One might then expect that if the observed amount of disorder and strength was greater than some critical value,
that the resulting random frustration would then first drive the system into a semi-classical spin glass state.
We are not aware of experimental studies having reported results suggesting a spin glass state in
Yb$_2$Ti$_2$O$_7$. 

Let us end by stating that further experimental studies of Yb-based pyrochlores to search for either the $U(1)$ 
liquid or the CFM phase are certainly most warranted. Perhaps variants such as
Yb$_2$Ge$_2$O$_7$, Yb$_2$Zr$_2$O$_7$ or Yb$_2$Hf$_2$O$_7$ could display interesting properties.
As discussed in Section 3, we would expect a material 
finding itself in the $U(1)$ quantum spin liquid state to be robust for a finite variation in the microscopic interaction parameters.
In this context, it may be worth noting that the disordered Yb$_2$GaSbO$_7$ material does not display
a sharp specific heat peak near $0.2$ K with its uniform susceptibility below
10 K  characterized by an {\it antiferromagnetic} Curie-Weiss temperature, $\theta_{\rm CW} \sim -2.3$ K,
unlike Yb$_2$Ti$_2$O$_7$ for which $\theta_{\rm CW}$ is ferromagnetic-like with 
$\theta_{\rm CW} \sim +0.4$ K  \cite{Blote1969}. 
Notwithstanding the random spin coupling caused by the
GaSb randomness on the transition metal ion site, a new generation of experiments (e.g AC and DC susceptibility, 
neutron, M\"ossbauer, muSR measurements) on Yb$_2$GaSbO$_7$ may prove interesting.

We conclude by saying that the three materials discussed in this section,
Tb$_2$Ti$_2$O$_7$, Pr$_2$(Sn,Zr)$_2$O$_7$ and Yb$_2$Ti$_2$O$_7$ are all described to some extent by the
type of effective spin-1/2 Hamiltonian of Eqn.~(\ref{eq:Heff}) from which one obtains an exotic gapless $U(1)$ spin liquid
state with gapped electric and magnetic excitations, in addition to gapless photons.
However, none of them have (yet) been found to be a clear realization of such a state.
The wide variety of materials in the family of R$_2$M$_2$O$_7$ pyrochlore oxides  \cite{Gardner_RMP}
offer the possibility that one or more of these compounds may eventually prove 
 compelling candidates to support
the exotic physics of the $U(1)$ spin liquid. In such a case, the experimental and theoretical lessons learned 
while investigating (Tb,Pr,Yb)$_2$(Ti,Zr,Sn)$_2$O$_7$ will undoubtedly prove useful.

It would also perhaps be interesting to investigate in more detail the AR$_{2}$S$_{4}$ and AR$_{2}$Se$_{4}$ (A$=$Cd,Mg) chalcogenide spinels in which the R$^{3+}$ rare-earth ion sits on a pyrochlore lattice \cite{Cd2Er2Se4,Lau2005}. In the context of spin ice-like systems, CdDy$_{2}$Se$_{4}$ likely has XY-like Dy$^{3+}$ ions and could prove particularly interesting \cite{Wong2013}.

\section{Conclusion}

In the foregoing sections, we have given an account of a theoretical proposal that effective spin-$1/2$
 pyrochlore magnets, close to the Ising limit, may have a spin liquid ground state with gapless photon-like excitations $-$
 the so-called quantum spin ice state. 
On the purely theoretical front, there is strong evidence both from effective field theory and numerics  that quantum fluctuations acting on the set of spin ice states can lead to just such a quantum spin liquid ground state. We can, moreover, make strong statements about the experimental observables in quantum spin ice. For example, unlike the case with many spin liquids, the gapless excitations carry spin one and couple directly to neutrons and would therefore be resolved as a sharp band of scattering intensity revealing the linear dispersion of the emergent photon.

Quantum spin ice exhibits a hierarchy of energy scales. The crossover between the quantum spin ice state and classical spin ice is controlled by the magnitude of the hexagonal ring exchange. At higher scales, there is a gap to the creation of spin ice defects $-$ the magnetic monopoles of Ref.~[\onlinecite{Castelnovo2008}]. The schematic recipe for making a quantum spin ice is that the material exchange should be dominated by an Ising coupling either coming purely from the balance of exchange parameters or assisted by the presence of an Ising-like single-ion anisotropy. There should be weaker transverse exchange couplings which generate the effective ring exchange. For realistic values of the couplings in real materials, the hexagonal ring exchange can vary over many orders of magnitude but, inevitably, in rare-earth magnets it will commonly be 10 to 100 times smaller than $J_\parallel$.
 If this is the case, then it is important to consider the effect of competing couplings on the quantum spin ice state. The success of the experimental search for quantum spin ice is contingent on whether the spin liquid is natural in the space of available couplings within an effective spin-$1/2$  model: which consists of four linearly independent nearest neighbour exchange, the long-range dipole coupling, further neighbour exchange, higher order ring exchange and, potentially, couplings to non-magnetic degrees of freedom. 
 
On the theoretical side, recent work has mapped out the phase diagram in the presence of the symmetry-allowed nearest neighbour exchange couplings using a type of gauge mean-field theory that allows one to study directly the fractionalized gauge theory degrees of freedom as well as conventional magnetically ordered phases \cite{Savary2011,SavaryFT}. From an experimental point of view, the finding from this study is encouraging: the quantum spin liquid ground state lives in a significant region in the space of parameters. It would be interesting to look at the situation beyond mean field theory and to consider further the phenomenology of quantum spin ice as it crosses over into classical spin ice. One issue is how one might probe directly the gapped excitations in the quantum spin ice state and their classical analogues at higher temperature.

On the experimental side, we have discussed in some detail four materials among the rare-earth pyrochlores which meet the simple criteria of having an Ising anisotropy with weaker transverse fluctuations of different microscopic origin. Unfortunately, while all four materials are associated with potentially very interesting open questions relating to the effect of disorder (Yb$_{2}$Ti$_{2}$O$_{7}$ and Tb$_{2}$Ti$_{2}$O$_{7}$) and the precise nature of the low temperature state (Tb$_{2}$Ti$_{2}$O$_{7}$,  Yb$_{2}$Ti$_{2}$O$_{7}$ and the Pr$^{3+}$ based materials), none appears to exhibit cleanly the quantum spin ice phase that is the main topic of this review. The existence of these materials does, however, demonstrate the naturalness of magnets in the vicinity of classical spin ice with added quantum fluctuations described, at least partially, by a model such as in Eqn.~(\ref{eq:Heff}). 
In addition, the last few years have seen considerable theoretical progress in  developing a quantitative
understanding of this class of materials. Given that many magnets among the pyrochlore rare-earths remain to be investigated - a number have been mentioned in the preceding pages - and given our, now, reasonable mature understanding of the broad series of pyrochlore magnets, the next few years should see similar rapid progress in mapping out the experimental parameter space for these materials. Optimistically, within a few years, we will understand that the materials discussed in this review are skirting around the edge of a real and significant region exhibiting spin liquid ground states and we will also have discovered examples of real pyrochlore materials that do live in that region and are the gapless spin liquids advertised in the title of this review.

\section{Acknowledgements}

We thank Zhihao Hao, Roderich Moessner, Frank Pollmann, Pedro Ribeiro and Arnab Sen for useful discussions.
This work is supported in part by  the NSERC of Canada and the
Canada Research Chair program (M.J.P.G., Tier 1) and by the Perimeter Institute for Theoretical Physics. Research at the Perimeter Institute is supported by the Government of Canada through Industry Canada and by the Province of Ontario through the Ministry of Economic Development \& Innovation.

\end{document}